\documentclass[preprint,showpacs,floatfix,amsmath,amssymb,superscriptaddress,aps,prb]{revtex4-2}
\usepackage[justification=raggedright]{caption}  

\usepackage{amsfonts}
\usepackage{stmaryrd}
\usepackage{bbm}
\usepackage{mathrsfs}
\usepackage{tipa}
\usepackage{amssymb}
\usepackage{txfonts}
\usepackage{graphicx}
\usepackage{dcolumn}
\usepackage{epstopdf}
\usepackage[colorlinks,linkcolor=blue,urlcolor=blue,citecolor=blue]{hyperref}
\usepackage{multirow}
\usepackage{subcaption}
\usepackage{booktabs} 
\usepackage{url}
\usepackage{upgreek}
\usepackage[utf8]{inputenc}
\usepackage[english]{babel}
\begin{document}

\title{Selective excitation of collective modes in multiband superconductor MgB\textsubscript{2}}

\author{Jiayu Yuan}
\affiliation{International Center for Quantum Materials, School of Physics, Peking University, Beijing 100871, China}

\author{Liyu Shi}
\affiliation{International Center for Quantum Materials, School of Physics, Peking University, Beijing 100871, China}

\author{Tiequan Xu}
\affiliation{Applied Superconductivity Center and State Key Laboratory for Mesoscopic Physics, School of Physics, Peking University, Beijing 100871, China}

\author{Yue Wang}
\affiliation{Applied Superconductivity Center and State Key Laboratory for Mesoscopic Physics, School of Physics, Peking University, Beijing 100871, China}

\author{Zizhao Gan}
\affiliation{Applied Superconductivity Center and State Key Laboratory for Mesoscopic Physics, School of Physics, Peking University, Beijing 100871, China}

\author{Hao Wang}
\affiliation{International Center for Quantum Materials, School of Physics, Peking University, Beijing 100871, China}

\author{Tianyi Wu}
\affiliation{International Center for Quantum Materials, School of Physics, Peking University, Beijing 100871, China}

\author{Dong Wu}
\affiliation{Beijing academy of quantum information science, Beijing, 100193, China}

\author{Tao Dong}
\email{taodong@pku.edu.cn}
\affiliation{International Center for Quantum Materials, School of Physics, Peking University, Beijing 100871, China}

\author{Nanlin Wang}
\email{nlwang@pku.edu.cn}
\affiliation{International Center for Quantum Materials, School of Physics, Peking University, Beijing 100871, China}
\affiliation{Beijing academy of quantum information science, Beijing, 100193, China}
\affiliation{Collaborative Innovation Center of Quantum Matter, Beijing, China}

\date{\today}

\begin{abstract}

Recent developments in nonequilibrium and nonlinear terahertz (THz) spectroscopies have significantly advanced our understanding of collective excitations in superconductors. However, there is still debate surrounding the identification of Higgs or Leggett modes, as well as BCS charge fluctuations, in the well-known two-band superconductor MgB$_2$. Here, we utilized both multi-cycle and single-cycle THz pump-broadband THz probe techniques to investigate the THz nonlinear response of MgB$_2$. Through multicycle THz pump-THz probe experiments on MgB$_2$, we observed distinct nonlinear signals at both the fundamental frequency ($\omega$) and the second harmonic frequency (2$\omega$) of the pump pulses, which exhibited resonant enhancement at temperatures where their frequencies respectively match 2$\Delta_{\pi}(T)$. They are mainly attributed to the $\pi$-band Higgs mode. By adjusting the THz pump pulse to a single-cycle waveform that satisfies non-adiabatic excitation criteria, we observed an over-damped oscillation corresponding to the Leggett mode. Our findings contribute to solving the ongoing debates and demonstrate the selective excitation of collective modes in multiband superconductors, offering new insights into the interaction between Higgs and Leggett modes. 

\end{abstract}

\maketitle

\textbf{Introduction}

Investigating collective modes in superconductors is a critical area of research in condensed matter physics because these modes provide deep insights into the underlying properties of superconducting states and their interactions. In a single band superconductor, the spontaneous breaking of continuous U(1) symmetry leads to the emergence of two important collective modes: a massless phase (or Nambu-Goldstone) mode and a massive 'Higgs' amplitude mode \cite{anderson_random-phase_1958,anderson_plasmons_1963,littlewoodAmplitudeCollectiveModes1982,pekker_amplitudehiggs_2015}. Due to the Anderson-Higgs mechanism, the phase mode is absorbed into the longitudinal component of the electromagnetic field, raising its energy to the plasma frequency. The Higgs mode, representing amplitude fluctuation of the complex order parameter, has the excitation energy at the superconducting gap (2$\Delta$) in the long-wavelength limit \cite{littlewoodAmplitudeCollectiveModes1982}. Detecting Higgs modes directly proves to be challenging due to their charge neutrality and spin inactivity, which requires nonlinear coupling to electromagnetic field without disturbing superconducting condensate. It is only in cases where superconductivity is accompanied by a charge density wave (CDW) order, like in 2H-NbSe$_2$, that the Higgs mode can be detected using Raman spectroscopy \cite{tsangRamanSpectroscopySoft1976,sooryakumarRamanScatteringSuperconductingGap1980,littlewoodAmplitudeCollectiveModes1982,meassonAmplitudeHiggsMode2014,pekker_amplitudehiggs_2015}. Recent advancements of intense ultrashort terahertz (THz) pulses have allowed observation of Higgs modes through two experimental protocols: free oscillations at $2\Delta$ after nonadiabatic excitation ($\tau_{p} \Delta < 1$) and third harmonic generation (THG) via periodic driving ($\tau_{p} \Delta \gg 1$) through a two-photon process, where $\tau_{p}$ is the pulse duration \cite{matsunaga_higgs_2013,matsunaga_light-induced_2014,PhysRevB.90.014515,PhysRevB.92.064508,shimano_higgs_2020,chu_phase-resolved_2020,wang_transient_2022,dongRecentDevelopmentUltrafast2023}. Despite progress in understanding nonlinear response of superconductivity, distinguishing Higgs modes from single-particle excitations (BCS charge fluctuations) remains challenging, as they share the same energy at long wavelengths and are influenced by sample disorder and band structure \cite{cea_nonlinear_2016,silaevNonlinearElectromagneticResponse2019a,tsujiHiggsmodeResonanceThird2020,seiboldThirdHarmonicGeneration2021}.

Multiband superconductors host more collective modes compared to single-band superconductors. In two-band superconductors, each band has a Higgs mode related to amplitude fluctuation. In the phase sector, besides the overall phase fluctuation which is pushed up to the plasma energy, a relative phase fluctuation between the two bands emerges, which is called Leggett mode \cite{leggett_number-phase_1966}. The energy scale of the Leggett mode depends on the two superconducting gaps and the interband and intraband coupling strengths. The interplay between this mode and the Higgs mode complicates the identification of non-equilibrium response sources in multi-band systems \cite{krull_coupling_2016}.

MgB$_2$ is a prototypical two-band s-wave superconductor with $\mathrm{T_c}$ = 30-40 K, which features two types of electronic bands crossing the Fermi level E$\mathrm{_F}$: the three-dimensional $\pi$ electron band and the quasi-two-dimensional $\sigma$ hole band. A number of spectroscopic measurements have shown two distinct isotropic superconductivity gaps within the $\pi$ and $\sigma$ bands, $\Delta_{\pi}$ = 0.4-0.7 THz and $\Delta_{\sigma}$ = 1.3-1.9 THz, respectively \cite{szabo_evidence_2001,xu_time-resolved_2003,blumberg_observation_2007}. The two gap characteristics thus establish MgB$_2$ as a platform to study the Higgs and Leggett modes. The Leggett mode is successfully identified in MgB$_2$ by Raman spectroscopy \cite{blumberg_observation_2007}. Recently, single-cycle THz pump-probe investigation with a center frequency of 1.4 THz in MgB$_2$ shows that the Leggett mode dominates the nonlinear response through the two-photon sum frequency process $\chi^{(3)}(\omega_{p},\omega_{p},\omega_{probe})$ \cite{giorgianni_leggett_2019}. This conclusion is further supported by wavelength-dependent narrowband THz driving--broadband THz probe measurement, in which no nonlinear response is observed at 0.57 THz driving (near $\Delta_{\pi}$ = 0.44 THz), and a nonlinear signal becomes detectable only when the driving pulse nearly fulfills the resonant condition of $2\omega_p=\omega_L \approx $ 2.5 THz \cite{giorgianni_leggett_2019}. However, a subsequent multi-cycle THz-driven THG experiment suggested that the contribution of the $\pi$-band Higgs mode, rather than the Leggett mode, predominated the nonlinear response, as evidenced by systematic $2\omega_p = 2\Delta(T)$ resonances with different driving frequencies \cite{kovalev_band-selective_2021}. Following the experimental investigations, a theoretical study supported the findings in the latter work that THz THG response in MgB$_2$ would show a resonance only for the lower gap \cite{haenelTimeresolvedOpticalConductivity2021}. Yet, another theoretical investigation indicated that the Leggett mode would dominate the nonlinear response when the pumping frequencies match one-half of its eigenfrequency, whereas the narrowband THz THG response was contributed by BCS charge fluctuations of the $\pi$-band gap rather than its Higgs mode \cite{fiore_contribution_2022}. The controversy then results in the following questions: First, how can we distinguish $\pi$-band Higgs mode from BCS charge fluctuations \cite{giorgianni_leggett_2019,kovalev_band-selective_2021,reinhoffer_high-order_2022}? Second, can we find the THz waveform that best selectively excites a given collective mode (Higgs mode of $\pi / \sigma$ band and Leggett mode) by tuning its spectrum and pulse duration? Developing an experimental protocol to identify the collective modes in such a conventional multi-band superconductor is also a crucial initial step in utilizing the technique for unconventional multi-band superconductors, such as iron-pnictide and nickel-based superconductors.

To address the above questions, in this work, we systematically investigate the nonlinear response of superconductivity collective modes in 10 nm MgB$_2$ thin films ($\mathrm{T_c}$=33 K, as shown in the Supplementary Information Fig.~S1) with comprehensive multi-cycle and single-cycle THz pump-broadband THz probe techniques. We initially use a multi-cycle THz pump to excite the MgB$_2$ sample. We observe that the $\pi$-band Higgs mode predominates the nonlinear response, as evidenced by both $2\omega_{p} = 2\Delta_{\pi}$ (T= 24 K) resonance in $\chi^{(3)}(\omega_{p},\omega_{p},\omega_{probe})$ and $\omega_{p} = 2\Delta_{\pi}$ (T=26 K) resonance in the $\chi^{(3)}(\omega_{p},\omega_{probe},-\omega_{probe})$ nonlinear channels after taking account of the screening effect. While the origin of the former resonance is still debated \cite{cea_nonlinear_2016,matsunaga_polarization-resolved_2017,haenelTimeresolvedOpticalConductivity2021,fiore_contribution_2022}, the latter resonance is primarily attributed to the Higgs mode of the superconductivity order parameter, rather than BCS fluctuations, as suggested by recent studies \cite{katsumiRevealingNovelAspects2024,katsumiAmplitudeModeMultigap2024}. In contrast, when we convert the multi-cycle pulse into a single-cycle waveform, we observe a strongly damped oscillation with a frequency of 1.8 THz $\pm$ 0.8 THz using single-cycle THz pump-probe spectroscopy in the non-adiabatic quench protocol. In comparison with reported experiments, we conclude that when the single-cycle pumping THz pulse has sufficient spectral overlap with the Leggett mode, the Leggett mode is activated and dominates the nonlinear response. These observations indicate that the Higgs and Leggett modes of MgB$_2$ can be selectively excited by tuning the spectrum and pulse duration of the THz pulses. 

\begin{figure*}[!ht]
  \centering
  \includegraphics[width=1.0\textwidth]{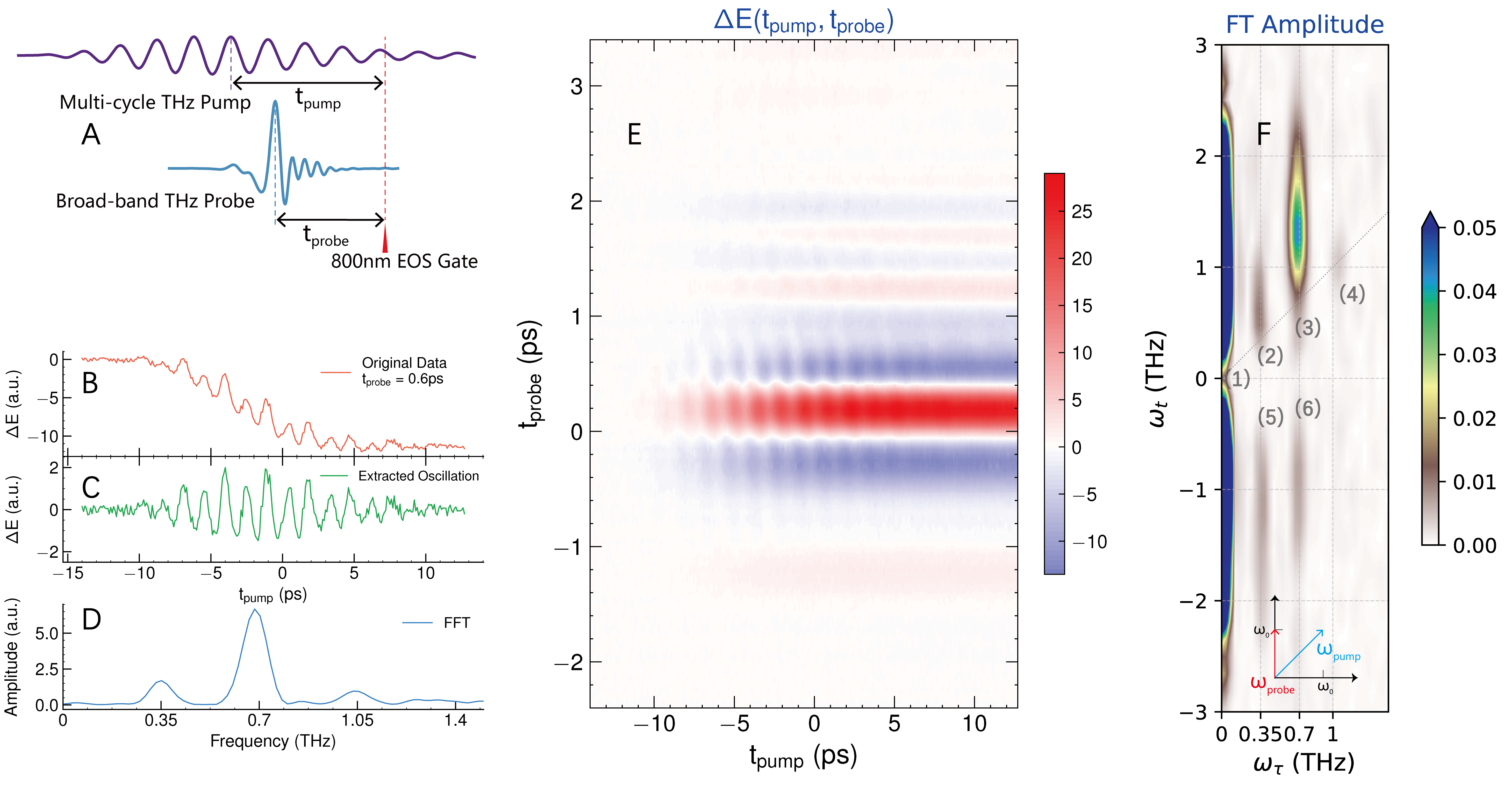} 
  \caption{0.35 THz multi-cycle pump-probe experiment results at 6 K.
  A: Definition of $\mathrm{t_{pump}}$ and $\mathrm{t_{probe}}$ in time-domain. B: Time-domain waveform of the nonlinear signal, cut at $\mathrm{t_{probe}}$ = 0.6 ps. C and D: Time- and frequency-domain waveforms of the extracted oscillation, cut at $\mathrm{t_{probe}}$ = 0.6 ps. E: Time-domain waveform of the nonlinear signal concerning $\mathrm{t_{pump}}$ and $\mathrm{t_{probe}}$. F: Frequency-domain waveform of the nonlinear signal $\Delta \mathrm{E} (\tau = {\mathrm{t_{pump}}} - {\mathrm{t_{probe}}}$).}
  \label{fig1}
\end{figure*}

\textbf{$\pi$-band Higgs mode response within a periodic driving}

Our study begins with multi-cycle THz pump-THz probe spectroscopy. The non-adiabatic excitation and the periodic driving of collective mode in superconductors are suggested to be equivalent \cite{matsunaga_light-induced_2014,udinaTheoryCoherentoscillationsGeneration2019}. However, there are conflicting experimental results for the two-band superconductor MgB$_2$. While previous research using both single-cycle THz quench and narrowband THz driving showed that the Leggett mode dominates the nonlinear response  \cite{giorgianni_leggett_2019}, but subsequent THz THG experiments suggest that the Higgs mode of the $\pi$-band is dominant \cite{kovalev_band-selective_2021}. To address this discrepancy, we conducted a systematic measurement using a narrowband THz pump and broadband THz probe on MgB$_2$ thin films (experimental setup details are provided in Supplementary Information III-VI). This systematic approach will also help distinguish between the $\pi$-band Higgs mode and its BCS fluctuation counterpart.  

Figure~\ref{fig1}A shows the time-domain waveforms of the narrowband pump pulse centered at 0.35 THz and broadband probe pulse, respectively. The pump-probe signal is recorded relative to the 800 nm gate pulse at time delays of $\mathrm{t_{pump}}$ and $\mathrm{t_{probe}}$ via electric-optical sampling (EOS). To visualize the oscillation and long-time dynamics in the superconducting state clearly, we plot the one-dimension pump-probe signal $\Delta \mathrm{E}(\mathrm{t_{pump}})$ as a function of $\mathrm{t_{pump}}$ in Fig.~\ref{fig1}B with setting $\mathrm{t_{probe}}$ to 0.6 ps, measured at 6 K. After THz driving, $\Delta \mathrm{E}(\mathrm{t_{pump}})$ gradually decreases with the time delay and reaches a plateau, whose value is proportional to the quasi-particle population arising from the pair breaking \cite{matsunaga_nonequilibrium_2012}. Besides the quasi-particle dynamics, clear oscillations are superposed on the transient curve. By fitting the background curve with an arctan function, we extract the pure oscillatory signal, which is shown in Fig.~\ref{fig1}C. Following the fast Fourier transformation (FFT), three distinct peaks are observed at twice the driving frequency $2\omega$ (0.7 THz), the fundamental frequency $\omega$ (0.35 THz), and three times the driving frequency $3\omega$ (1.05 THz), as shown in Fig.~\ref{fig1}D. The $2\omega$ peak is linked to the nonlinear process $\chi^{(3)}(\omega_{p},\omega_{p},\omega_{probe})$, the same nonlinear kernel observed in THz-THG experiments \cite{kovalev_band-selective_2021}. The origin of this nonlinear kernel is under debate, as mentioned above. The $3\omega$ peak is linked to nonlinear process $\chi^{(5)}(\omega_{p},\omega_{p},\omega_{p},\omega_{probe},-\omega_{probe})$. The appearance of the $\omega$ peak is unexpected, as it is thought to be symmetry-forbidden unless the inversion symmetry of the system is broken for some reasons \cite{yang_lightwave-driven_2019, nakamuraNonreciprocalTerahertzSecondHarmonic2020, vaswaniTerahertzSecondHarmonicGeneration2020, nakamuraPicosecondTrajectoryTwoDimensional2024a}. We then perform a second harmonic generation (SHG) on the investigated sample with 0.35 THz driving, yet no SHG signal is detected, ruling out steady inversion symmetry breaking as the cause of the $\omega$ peak. Previous experimental and theoretical studies indicate that when the driving electric field exceeds a critical field and reaches the non-perturbative regime, the THz lightwave accelerated supercurrent is able to break the inversion symmetry dynamically, resulting in a $\omega$ peak in the one-dimension pump-probe spectra \cite{yang_lightwave-driven_2019,mootzVisualizationQuantumControl2022}. To make sure that our pump electric field stays within the perturbative regime, we conducted a systematic fluence-dependent measurement as shown in Supplementary Information Fig.~S9. All THz driving measurements performed were within this perturbative regime.

To further identify the origin of this $\omega$ peak, we turn to the two dimension (2D) frequency space of the measured 2D $\Delta \mathrm{E}(\mathrm{t_{pump}},\mathrm{t_{probe}})$ signals, in which the different nonlinear process can be distinguished through the frequency wavevector analysis. The corresponding 2D-THz spectra are obtained by FFT of $\Delta \mathrm{E}(\mathrm{t_{pump}},\mathrm{t_{probe}})$ with respect to both $t=\mathrm{t_{probe}}$ (frequency $\omega_t$) and $\tau= \mathrm{t_{pump}}-\mathrm{t_{probe}}$ (frequency $\omega_{\tau}$), where $\tau$ represents the pump-probe relative time (the definitions of $t$ and $\tau$ are detailed in Supplementary Information Fig.~S7), as shown in Figure~\ref{fig1}F. The frequency vectors of $\omega_{\tau}$ and $\omega_t$ are $\omega_{\tau} = (\omega_p, \omega_p)$ and $\omega_t = (0, \omega_0 \pm \Delta \omega_0 / 2)$, where $\omega_0 = 1.0 \, \text{THz}$ is the center frequency of the broadband probe pulse, and $\Delta \omega_0 = 1.3 \, \text{THz}$ denotes the broadband probe frequency width (the THz probe spectrum is detailed in Supplementary Information Fig.~S6). In the two-dimensional frequency spectrum, six distinct peaks are observed with varying degrees of broadening along the $\omega_t$ axis. These peaks are positioned at specific frequencies along the $\omega_\tau$ axis: (1) 0 THz, (2) and (5) $\omega_p$=0.35 THz, (3) and (6) 2$\omega_p$=0.7 THz, and (4) 3$\omega_p$=1.05 THz. The generation of these peaks can be attributed to distinct nonlinear processes. The pump-probe processes $(\omega_{p}, -\omega_{p}, \omega_{probe})$ and $(\omega_{probe}, -\omega_{probe}, \omega_{p})$ generate peaks at $(0, \pm (\omega_0 \pm \Delta \omega_0 / 2) )$ and $(\omega_p, \omega_p)$, respectively. The peak at (2$\omega_p$, 2$\omega_p$+$\omega_0 \pm \Delta \omega_0 / 2 $) is generated by $(\omega_{p},\omega_{p},\omega_{probe})$ process, which produces the THG signal in previous THz driving experiments. Interestingly, a peak at $(3\omega_p, 3\omega_p)$ is observed, generated by a fifth-order nonlinear process $(\omega_p, \omega_p, \omega_p, \omega_{probe}, -\omega_{probe})$, as the equilibrium THG signal has been subtracted. Additionally, two peaks appear in the region where $\omega_t$ is less than zero. For peak (5), both the rephasing process $(\omega_{p}, -\omega_{probe}, -\omega_{probe})$ and the two-quantum process $(-\omega_{probe}, -\omega_{probe}, \omega_{p})$ contribute to the same feature at ($\omega_p$, $\omega_p$ - 2$(\omega_0 \pm \Delta \omega_0 / 2)$). Similarly, for peak (6), the rephasing process $(-\omega_{probe}, \omega_{p}, \omega_{p})$ and the two-quantum process $(\omega_{p}, \omega_{p}, -\omega_{probe})$ contribute to a single feature at ($2\omega_p$, $2\omega_p$ - $(\omega_0 \pm \Delta \omega_0 / 2)$). One might question why the peak intensity at position (2) appears weaker than at position (3) while being similar to position (5). This is unexpected, as the four-wave mixing pump-probe signals (1) and (2) are generally anticipated to be significantly stronger than the THG signal (3) and the rephasing or two-quantum signal (5). We attribute this phenomenon to a key factor: in our experimental setup, the pump and probe polarizations are perpendicular to each other. Additionally, a polarizer is positioned after the sample, aligned perpendicularly to the pump direction, effectively suppressing the primary component of the transmitted pump. This significantly weakens the nonlinear signal $\Delta \mathrm{E}(\omega_{probe}, -\omega_{probe}, \omega_{p})$, which is aligned with the pump polarization. In contrast, the THG signal $\Delta \mathrm{E} (\omega_{p}, \omega_{p}, \omega_{probe})$ and the rephasing or two-quantum signal $\Delta \mathrm{E} (\omega_{p}, -\omega_{probe}, -\omega_{probe})$ are less affected, as their polarizations lie in an intermediate orientation between the pump and probe. Taking this factor into account, we can understand why the peak at position (2) is weaker than that at position (3) and comparable to the peak at position (5) in the 2D spectrum.

The $(\omega_{p}, \omega_{p})$ and $(\omega_{p}, \omega_{p} - 2(\omega_{0} \pm \Delta \omega_{0} / 2))$ peaks contribute to the $\omega$ oscillation, while the $(2\omega_{p}, 2\omega_{p} + (\omega_{0} \pm \Delta \omega_{0} / 2))$ and $(2\omega_{p}, 2\omega_{p} - (\omega_{0} \pm \Delta \omega_{0} / 2))$ peaks generate the $2\omega$ oscillation in the one-dimensional pump-probe spectra. As elaborated below, the $\omega$ signal provides an alternative perspective for identifying the primary contributions to the nonlinear response in superconductors.

\begin{figure*}[!ht]
  \centering
  \includegraphics[width=1.0\textwidth]{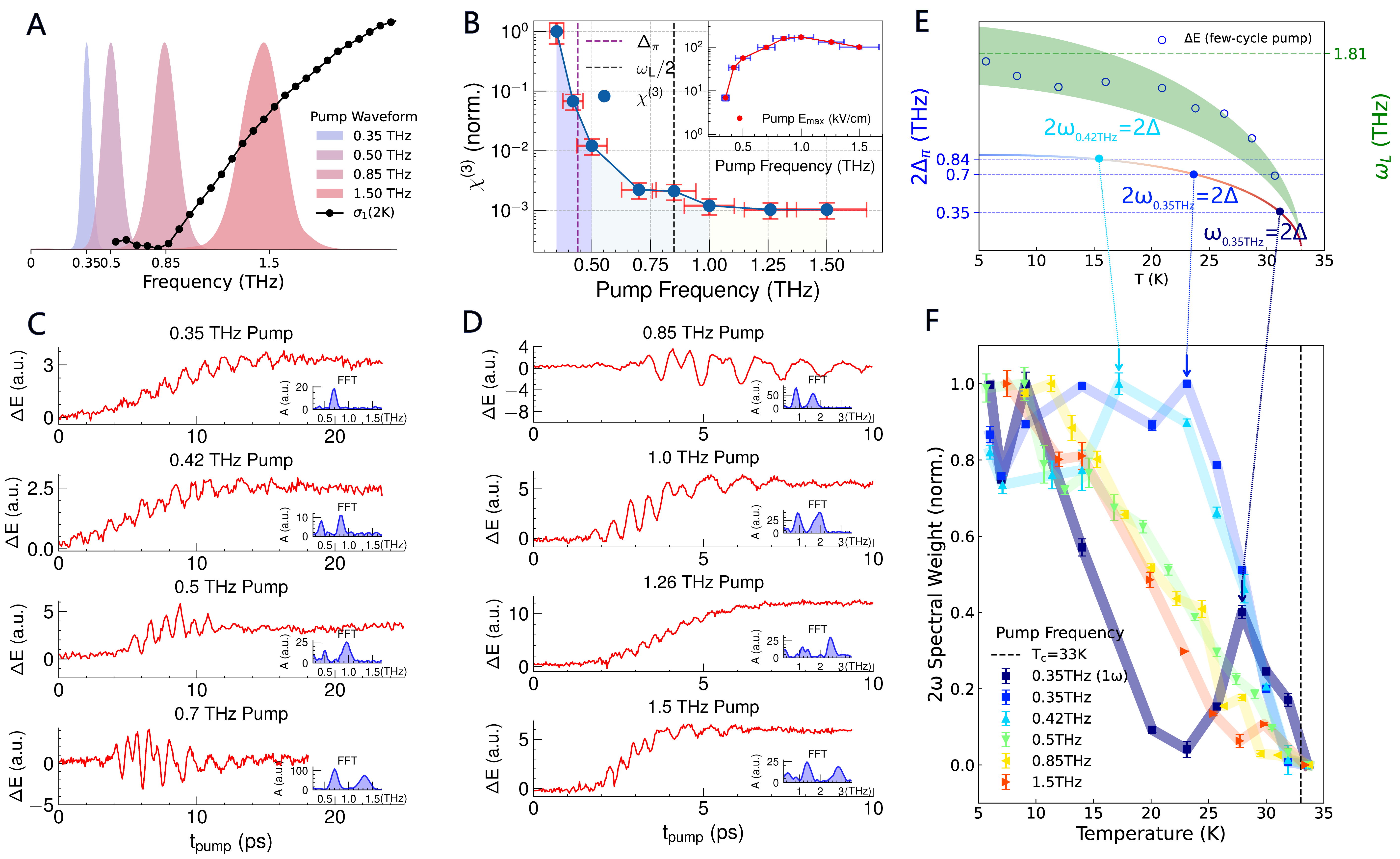} 
  \caption{The nonlinear signal from multi-cycle pump-probe measurements at different pump frequencies is presented. A: The spectral waveform of the multi-cycle pump, with the black curve representing the THz real conductivity. B: The nonlinear coefficient $\chi^{(3)}$ plotted against pump frequency, with the inset showing the pump's peak electric field. C and D: Time-domain and frequency-domain waveforms of the nonlinear signal at various pump frequencies. E: The temperature dependence of 2$\Delta_{\pi}$ and $\omega_L$, with horizontal dotted lines marking $\omega$=0.35 THz and 2$\omega$ = 0.7, 0.84 THz. F: Normalized $2\omega$ (or $1\omega$) spectral weight as a function of temperature at different pump frequencies.}
  \label{fig2}
\end{figure*}

Having identified the nonlinear process of $\omega$ and 2$\omega$ oscillations, we then perform the 1D pump-probe measurement with a given $\mathrm{t_{probe}}$ = 0.2 ps under various driving frequencies. The spectra of the narrowband driving pulses and $\sigma_1(\omega, T=2\,\text{K})$ of MgB$_2$ are shown in Fig.~\ref{fig2}A (the calculation of $\sigma_1$ from equilibrium THz spectroscopy is detailed in Supplementary Information Fig.~S2 and S3). At 6 K, clear $2\omega$ and $\omega$ oscillations in $\Delta \mathrm{E}(\mathrm{t_{pump}})$ are observed at all driving frequencies (from 0.35 to 1.5 THz) with the selected fluences, as depicted in Fig.~\ref{fig2}B and C. We will discuss the $\omega$ signal later. It is worth noting that distinct $2\omega$ oscillations are visible when the pump frequency significantly undershoots the one-half of Leggett mode characteristic frequency ($\omega_L$/2=0.85 THz). This result differs from the previous experimental studies on MgB$_2$ \cite{giorgianni_leggett_2019}, where the THz-driven $2\omega$ oscillations appear unless the driving frequencies are close to one-half of Leggett mode frequency. Because of the superconducting condensation, the transmission coefficient of the driving THz pulses will change dramatically at 6 K when their frequencies cross the superconductivity gap (see the transmission curves in Supplementary Information Fig.~S10), e.g., $2\Delta_{\pi}=0.88$ THz. The third-order nonlinear susceptibility $\chi^{(3)} (\omega, T=6\,\text{K})$ is calculated using the formula ${\chi^{(3)} (\omega, T=6\,\text{K}) \propto\frac{\mathrm{E}(2\omega, T=6\,\text{K})}{[\mathrm{E_{inc}}(\omega)\mathrm{Tr}(\omega, T=6\,\text{K})]^2\mathrm{E_{probe}}(\omega_{probe})\mathrm{Tr}(\omega_{probe}, T=6\,\text{K})}}$. Since the field intensity and transmission coefficient of the THz probe pulse used for different multicycle driving pulse frequencies are the same, the factors from the THz probe pulse do not need to be considered when comparing $\chi^{(3)}$ with different driving frequencies. Therefore, we calculate the third-order nonlinear susceptibility $\chi^{(3)} (\omega, T=6\,\text{K})$ for all driving frequencies using the formula: ${\chi^{(3)} (\omega, T=6\,\text{K}) \propto\frac{\mathrm{E}(2\omega, T=6\,\text{K})}{[\mathrm{E_{inc}}(\omega)\mathrm{Tr}(\omega, T=6 \,\text{K})]^2}}$, where $\mathrm{E_{inc}}(\omega)$ represents the electric field of the incident THz pulse at the sample position, as shown in the inset of Fig.~\ref{fig2}B, while $\mathrm{Tr}(\omega, T=6\,\text{K})$ and $\mathrm{E}(2\omega, T=6 \,\text{K})$ correspond to the transmission coefficient of the driving pulse and the intensity of the recorded $2\omega$ oscillations at 6 K, respectively (details of the calibrations can be found in the Supplementary Information Table~S1-S3). After considering the screening effect, no resonance is still observed near driving frequencies where the Leggett mode's resonance condition would be fulfilled. This observation suggests that within narrowband THz driving protocol, the Leggett mode unlikely dominates the nonlinear response. Thus, we tend to ascribe the observed $2\omega$ signal to the amplitude fluctuations of the $\pi$-band superconductivity gap, as reported in previous THz-THG measurements \cite{kovalev_band-selective_2021}. 

We further study the temperature evolution of the oscillations to uncover their connections with superconductivity collective modes. To quantitatively evaluate the temperature dependence of $2\omega$ and $1\omega$ oscillations, we calculate their normalized spectral weights as $\mathrm{SW}^{\mathrm{norm}}_{2\omega}(T)=\frac{\mathrm{SW}_{2\omega}(T)}{\mathrm{Tr}^2(\omega,T)\mathrm{Tr}(\omega_{probe},T)}$ and $\mathrm{SW}^{\mathrm{norm}}_{1\omega}(T)=\frac{\mathrm{SW}_{1\omega}(T)}{\mathrm{Tr}(\omega,T)\mathrm{Tr}^2(\omega_{probe},T)}$, respectively, where $\mathrm{Tr}(\omega, T)$ and $\mathrm{Tr}(\omega_{probe}, T)$ denote the transmission coefficients of the pump and probe electric fields (see Supplementary Information Fig.~S10 and S12 for details). The results of $2\omega$ spectral weight below $\mathrm{T_{c}}$ and with all driving frequencies are summarized in Fig.~\ref{fig2}F. Weak resonances occurring at 2$\omega_p$= 2$\Delta_{\pi}(T)$, as indicated by the vertical dashed line in Fig.~\ref{fig2}E and Fig.~\ref{fig2}F, for 0.35 and 0.42 THz driving are visible, which is consistent with previous THz THG measurement \cite{kovalev_band-selective_2021}. This result further confirms our assignment of the $\pi$-band relevance being responsible for the observed $2\omega$ signal. The raw data, without considering the screening effect, is presented in Supplementary Information Fig.~S11, where the resonance behavior is more pronounced. Since whether the $\pi$-band Higgs mode or BCS fluctuations dominate the observed  $2\omega$ oscillations in MgB$_2$ currently is still under debate, we turn to the $\omega$ signal arising from the $(\omega_{p},\omega_{probe},-\omega_{probe})$ nonlinear process, which recently has been assigned to mainly contribute by amplitude mode of superconductivity order parameter through theoretical study and two dimensions THz spectroscopy \cite{katsumiRevealingNovelAspects2024}. Interestingly, we find that the temperature evolution of the $\omega$ signal is not trivial. Starting from 6 K, its intensity first decreases with increasing temperature until 23 K, which then is followed by a sharp increase towards 26 K. With further increasing temperature, the $\omega$ intensity continuously decreases towards $\mathrm{T_c}$. The $\omega$ intensity presents a resonant enhancement when the driving frequency matches the $\pi$-band superconductivity gap ($\omega_{p} = 2\Delta_{\pi}(T)$), as indicated by the dashed line in Fig.~\ref{fig2}E and Fig.~\ref{fig2}F. Moreover, the resonant behaviors presented in 2$\omega$ and $\omega$ channels are neither observed near temperatures where the resonance condition would be fulfilled for the $\sigma$-band gap, nor for the Leggett mode (see Fig.~\ref{fig2}E). With these observations, we suggest that the $\pi$-band Higgs mode predominates the nonlinear response under the narrow band driving condition, being consistent with the theoretical studies that predicted an intermediate disorder in superconductors, for example, the dirty limit $\pi$ band gap in MgB$_2$, would enhance the Higgs mode contributed paramagnetic fluctuations \cite{haenelTimeresolvedOpticalConductivity2021}. It is important to note that the predominance of $\pi$ band Higgs mode does not necessarily imply the absence of contribution from Leggett mode in the driving condition. In particular, for the driving frequencies $\omega_p > 2\Delta_{\pi}$, the contribution from Legget mode might be sizable, however, mixed with $\pi$ band Higgs mode, making them indistinguishable from each other. 

\textbf{Leggett mode response within non-adiabatic excitation}

\begin{figure}[!ht]
  \centering
  \includegraphics[width=\textwidth]{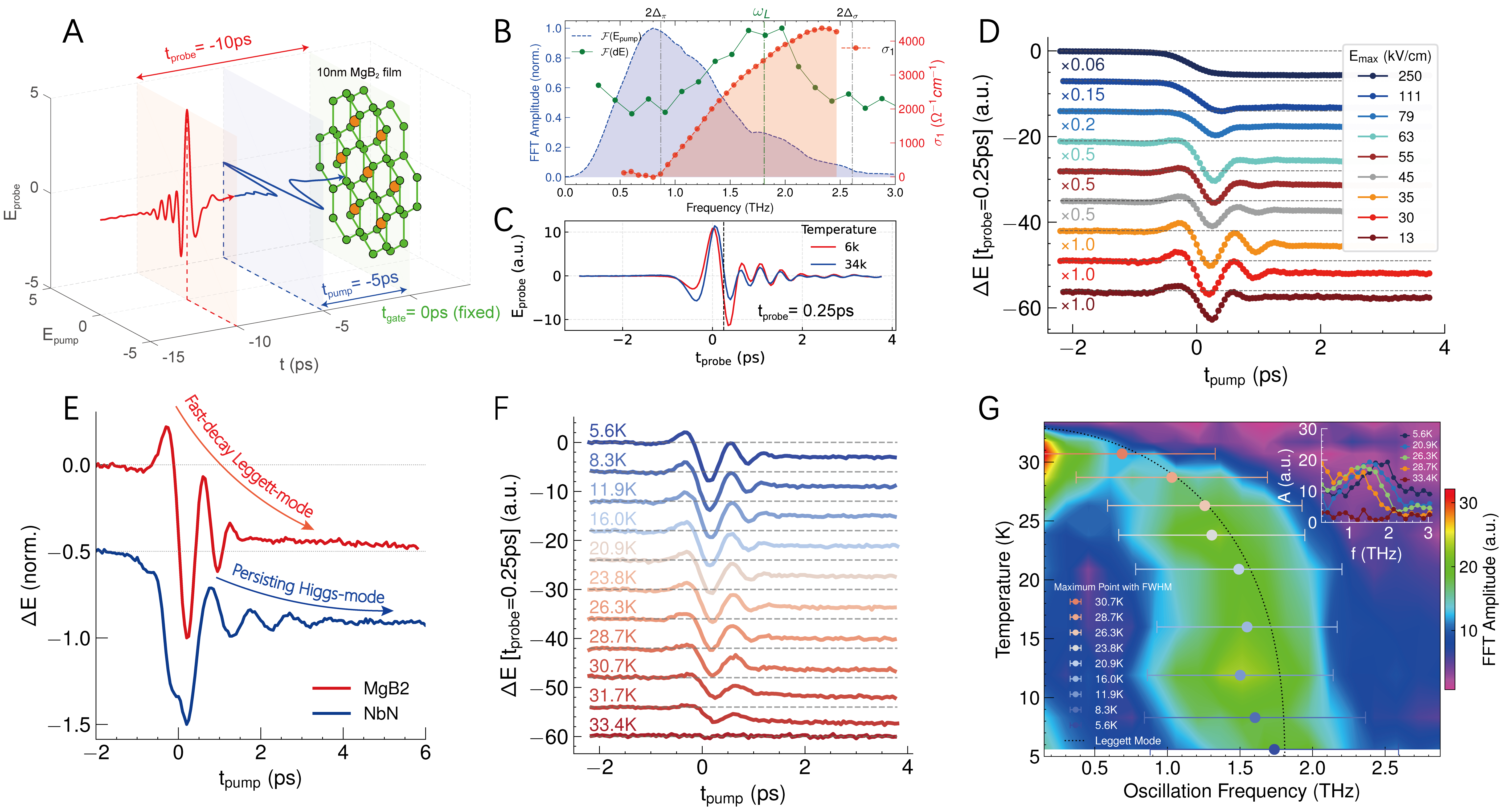} 
  \caption{THz single-cycle pump-probe experiment set-up and results. (A) Schematic of the experimental set-up. (B) The blue shaded area represents the THz pump's FT amplitude spectrum considering SC's screening effect, the green shaded area represents THz nonlinear signal's FT amplitude spectrum with dashed lines marked characteristic frequencies $\mathrm{\omega_L}$=1.8 THz and the red shaded area represents the real part conductivity $\sigma_1$ of MgB$_2$ at 2 K with dashed lines marked characteristic frequencies $2\Delta_\pi$=0.88 THz. (C) The relative position between gate and probe pulses. (D) Time domain waveform of nonlinear signal $\Delta \mathrm{E_{probe}}$ with varied pump fluence, measured at 6 K. (E) Comparison of nonlinear oscillation signals in MgB$_2$ and NbN under the same THz single-cycle pump-probe experimental conditions. (F) and (G) Time and frequency domain waveform of nonlinear signal $\Delta \mathrm{E_{probe}}$ at varied temperatures with a maximum pump electric field of 35 kV/cm, featuring a marked dashed line to show the evolution of the Leggett mode.}
  \label{fig3}
\end{figure}

To verify the selective excitation of collective mode responses in MgB$_2$ with different pump pulse durations, we replace the multi-cycle pump with a single-cycle pump and investigate whether the nonlinear response originates from the Leggett mode, as reported previously \cite{giorgianni_leggett_2019}. Figure~\ref{fig3}A shows the schematic diagram of our setup, where two THz pulses, with perpendicular electric field polarizations, are incident co-linearly on the c-axis oriented MgB$_2$ thin film. The transmitted nonlinear THz signal, denoted as $\Delta \mathrm{E_{probe}}(\mathrm{t_{pump}}, \mathrm{t_{probe}})$, is recorded using the double modulation technique (see Supplementary Information Fig.~S8 for details). The blue-shaded curve in Figure~\ref{fig3}B shows the spectrum of the single-cycle THz pumping pulse, with a peak at 0.8 THz and its corresponding duration is $\tau_p \sim$ 1 ps. The real part of the conductivity, $\sigma_1$, calculated from the linear THz transmission spectrum for MgB$_2$ at 2 K, is shown as the red-shaded curve in Fig.~\ref{fig3}B. Its onset absorption around 0.88 THz $\pm$ 0.05 THz is corresponding to the $\pi$-band superconductivity gap 2$\Delta_{\pi}$. We estimate $\sigma$-band superconductivity gap $\Delta_{\sigma}$ by a empirical formula $ \Delta_{\sigma}\sim 3 \Delta_{\pi}$, which then yields $2 \Delta_{\sigma} \sim 2.64 $ THz. With the two superconductivity gap values, we then estimate the eigenfrequency of the Leggett mode to be approximately $\omega_{L} \sim $ 1.8 THz (detailed in Supplementary Information Fig.~S14). These estimated values are marked by dashed lines in Fig.~\ref{fig3}B. Furthermore, these evaluated values give the ratios of the pump duration over the mode energy as: $\tau_{p} \Delta_{\pi} \sim 0.44$, $\tau_{p} \frac{\omega_L}{2} \sim 0.88$, and $\tau_{p} \Delta_{\sigma} \sim 1.32$, indicating that the pump pulse satisfies the nonadiabatic condition ($\tau_{p} \Delta < 1$) for both $\pi$-band Higgs mode and the Leggett mode \cite{papenkort_coherent_2007,matsunaga_higgs_2013}, which then enables one to expect simultaneous observation of $\pi$-band Higgs mode and Leggett mode free oscillations in MgB$_2$. 

Figure~\ref{fig3}C shows the terahertz waveforms of probe pulse at temperatures above and below $\mathrm{T_c}$. The reduced signal intensity below $\mathrm{T_c}$ is related to the superconducting condensation. Figuree~\ref{fig3}D shows the temporal evolution of $\Delta \mathrm{E_{probe}}(\mathrm{t_{pump}},\mathrm{t_{probe}}=0.25 \,\mathrm{ps})$ as a function of $\mathrm{t_{pump}}$ at different pump electric fields. At a high electric field with 250 kV/cm when the superconductivity is strongly excited, $\Delta \mathrm{E_{probe}}(\mathrm{t_{pump}},\mathrm{t_{probe}}=0.25 \,\mathrm{ps})$ shows a step function behavior without oscillation, being similar to previous study on NbN \cite{matsunaga_nonequilibrium_2012,matsunaga_higgs_2013}. As the pump electric field decreases, a heavily damped oscillatory signal gradually emerges, reaching its maximum at 35 kV/cm. With further decreasing the pumping electric field down to 13 kV/cm, the oscillation is undetectable (the comparison of $\mathrm{E^2_{pump}}$ and the $\Delta \mathrm{E}$ behavior to exclude a simple $\mathrm{E^2_{pump}}$-dependent nonlinear response is shown in Supplementary Information Fig.~S15). It deserves to be noted that the damping of oscillation at the optimal pumping electric field here is much faster than that observed in NbN, as displayed in Fig.~\ref{fig3}E, a feature that we shall address further below. After fixing the pump at optimal field $\sim$35kV/cm, we examine the temperature evolution of the oscillations. As shown in Fig.~\ref{fig3}F, the step-like pump-probe signal $\Delta \mathrm{E_{probe}}(\mathrm{t_{pump}},\mathrm{t_{probe}}=0.25 \,\mathrm{ps})$ and the accompanied overdamped oscillation gradually diminish as temperature increases. Only when the temperature exceeds $\mathrm{T_c}$, both the step-like long-lived dynamics and oscillation in $\Delta \mathrm{E_{probe}}(\mathrm{t_{pump}},\mathrm{t_{probe}}=0.25 \,\mathrm{ps})$ disappear completely, signaling their superconductivity origin. To extract the frequency of the oscillations, we analyze the oscillating segments of the $\Delta \mathrm{E_{probe}}(\mathrm{t_{pump}},\mathrm{t_{probe}}=0.25 \,\mathrm{ps})$  after $\mathrm{t_{pump}} =$ 0.2 ps at different temperatures using segmented fast Fourier transformation (FFT). Figure~\ref{fig3}G shows the FFT amplitude spectra against temperatures in a two-dimensional color map and the oscillation frequency at various temperatures. The 1D FFT spectrum at various temperatures is shown in the inset of Fig.~\ref{fig3}G, and its center frequency and error bar are determined by Lorentz fitting of the FFT spectra. A broad peak with a central frequency of 1.8 THz is observed, and it softens with increasing temperature. This frequency matches the eigenfrequency of the Leggett mode $\omega_{L}$ of MgB$_2$, as indicated by a dashed line in Fig.~\ref{fig3}B. Therefore, this result strongly suggests the observed damped oscillation arises from Leggett mode rather than the $\pi$-band Higgs mode after a nonadiabatic excitation, which is consistent with the reported Raman spectroscopy \cite{blumberg_observation_2007} and single-cycle THz pump-probe measurement \cite{giorgianni_leggett_2019}. Since Leggett mode does not have a competing counterpart like BCS fluctuations for Higgs mode in the nonlinear response, the observed overdamped oscillation can confidently be ascribed to a coherent motion of the superconductivity order parameter, e.g., Leggett mode. 

To further verify the observed damped oscillation in MgB$_2$ arising from the intrinsic nonlinear response of the collective modes rather than the insufficient resolution of our spectrometer, we repeated the single-cycle pump-probe experiments on NbN ($\mathrm{T_c}$ = 13 K, 2$\Delta \sim$ 1.1 THz). For the NbN sample, as shown in Fig.~\ref{fig3}E, we observe a long-lived free oscillation with a frequency of 1.1 THz in $\Delta \mathrm{E_{probe}}(\mathrm{t_{pump}},\mathrm{t_{probe}}=0.25\,\mathrm{ps})$, which is a characteristic feature of Higgs mode of the superconductivity order parameter in a single-band s-wave superconductor evidenced in Ref. \cite{matsunaga_higgs_2013}. Conversely, the temporal evolution of $\Delta \mathrm{E_{probe}}(\mathrm{t_{pump}},\mathrm{t_{probe}}=0.25\,\mathrm{ps})$ of MgB$_2$ shows an overdamped oscillation with a frequency of  1.8 THz $\pm$ 0.8 THz and the expected free oscillation leading by $\pi$-band Higgs mode is invisible even our pumping pulse fulfills the non-adiabatic excitation condition ($\omega_p \sim $0.7 THz,$\tau_{p} \Delta_{\pi} \sim 0.44 $) for $\pi$-band Higgs mode. This comparison thus indicates that the presence of inter-band coupling between the two superconductivity order parameters changes the spectrum of collective modes and affects their nonlinear responses. A simple extension of experimental protocols used in the single-band superconductor to the two-band case is not sufficient to resolve the relevant collective modes. It is worth noting that our pumping THz pulse has a lower frequency spectra component around $2\Delta_{\pi} \sim 0.88 $ THz and its tail extends to high frequency while the THz pulse in Ref. \cite{giorgianni_leggett_2019} has more spectra weight at one half of the Leggett mode energy, both pumping THz pulses satisfies the non-adiabatic excitation condition for $\pi$-band Higgs and the Leggett modes. The absence of $\pi$-band Higgs mode contributed free oscillation and the presence of Leggett mode induced oscillation in two cases suggest that once the pumping THz pulse is in the non-adiabatic excitation condition for Leggett mode of MgB$_2$, Leggett mode will be coherently activated and dominate the nonlinear response.

\textbf{Summary}

In summary, we have demonstrated selective excitation of collective modes in a prototype multiband superconductor MgB$_2$ utilizing different schemes of THz pump-probe spectroscopy. By tuning the excitation wavelengths and pulse duration (or bandwidth) of the pumping THz pulse, we can selectively excite and probe the $\pi$ band Higgs mode under a periodic driving and Leggett mode in a non-adiabatic quenching condition. The dichotomous nonlinear response in MgB$_2$ after periodic driving and non-adiabatic excitation indicates that the inter-band coupling induced Leggett mode sets an upper boundary of the THz pumping spectrum for the observation of the free oscillation of $\pi$ band Higgs mode. A fine-tuning of wavelengths and durations of the excitation THz pulse in a single-cycle waveform without sufficient overlap with Leggett mode is highly desirable to observe the expectant $\pi$ band Higgs mode free oscillation. Our observations suggest that a full analysis of different nonlinear processes provides a unique route for distinguishing the Higgs mode from its BCS fluctuations counterpart in the nonlinear response of superconductors. Moreover, our demonstration of selective excitation of corresponding collective modes will be instrumental in studying and controlling the coupling between Higgs mode and Leggett mode in two strong electric field THz pulses measurement, e.g., one pulse in the non-adiabatic excitation condition for Leggett mode and the second one in the periodic driving of $\pi$ band Higgs mode.

\begin{acknowledgments}

We gratefully acknowledge L. Benfatto and Y. Wan for illuminating discussions. This work was supported by the National Natural Science Foundation of China (No.12488201, 12250008), and the National Key Research and Development Program of China (2021YFA1400200, 2022YFA1403901). 

\end{acknowledgments}

\begin{center}
    \textbf{\large Supplementary Information}  
\end{center}
\setcounter{figure}{0}

\section{Sample characterization}

\begin{figure}[!h]
    \centering
        \renewcommand{\figurename}{Fig S}

    \includegraphics[width=0.5\linewidth]{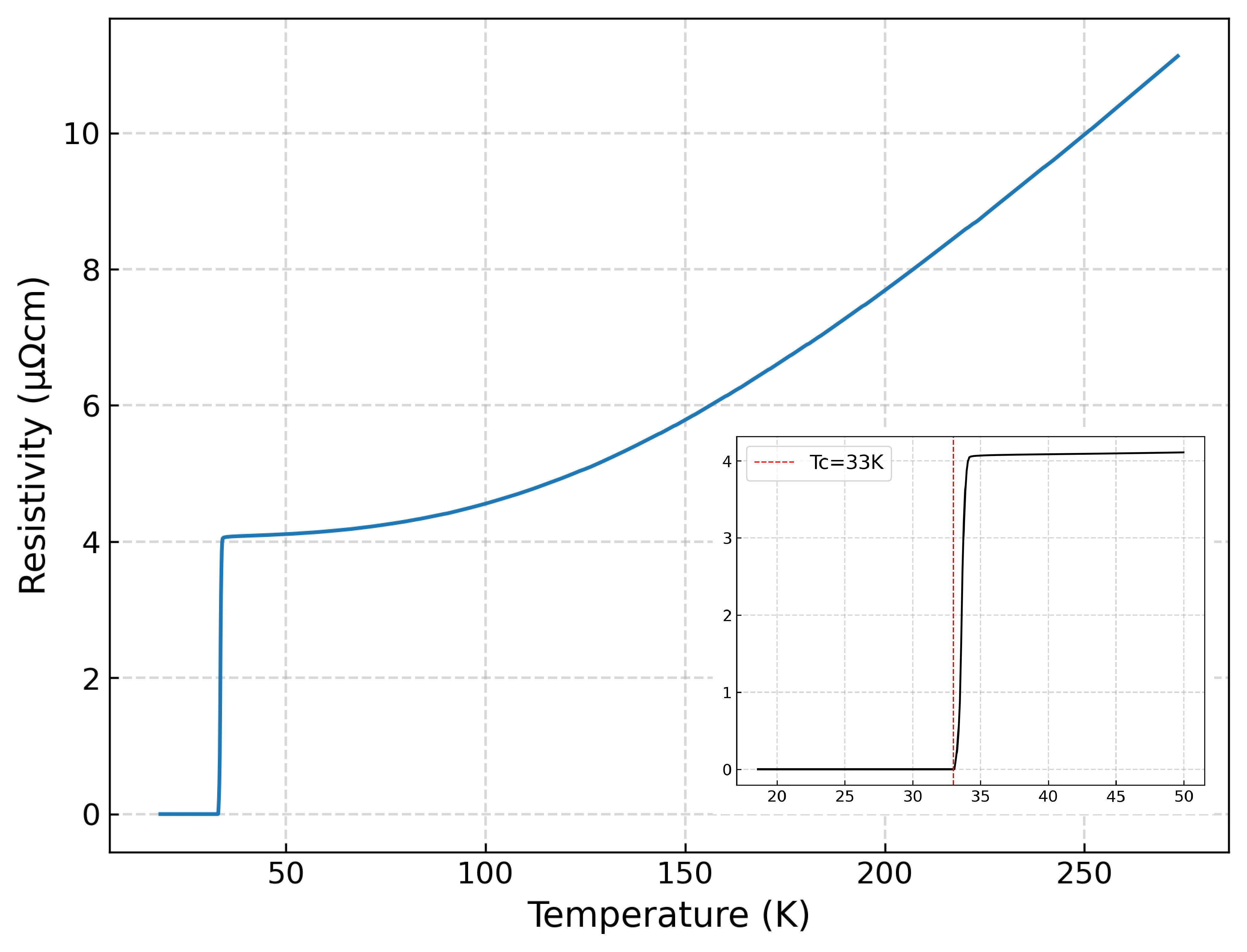}
    \caption{The resistivity characterization of the measured thin films.}
    \label{S1}
\end{figure}

The MgB$_2$ thin film, prepared via pulsed laser deposition (PLD), is 10 nm thick and deposited on a 0.5 mm thick MgO substrate. Fig.~S\ref{S1} presents the temperature-dependent in-plane resistivity of the MgB$_2$ film.

\section{Optical conductivity of the MgB$_2$ thin film}

\begin{figure}[!h]
    \centering
        \renewcommand{\figurename}{Fig S}

    \includegraphics[width=0.8\linewidth]{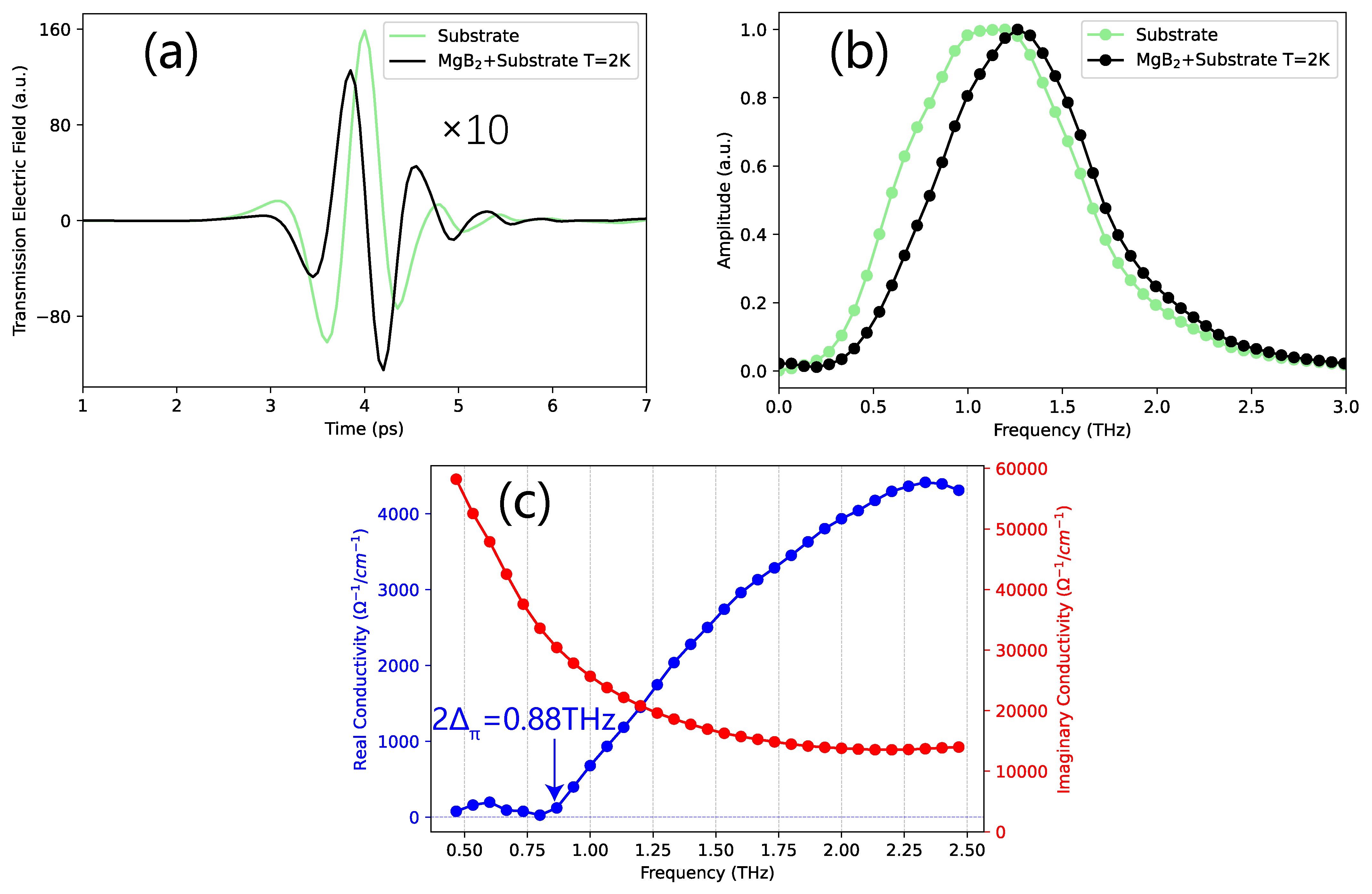}
    \caption{(a) Transmission THz time-domain spectrum, (b) frequency-domain spectrum, and (c) calculated optical conductivity of MgB$_2$ at $2 \, \text{K}$.}
    \label{S2}
\end{figure}

\begin{figure}[!h]
    \centering
        \renewcommand{\figurename}{Fig S}

    \includegraphics[width=0.8\linewidth]{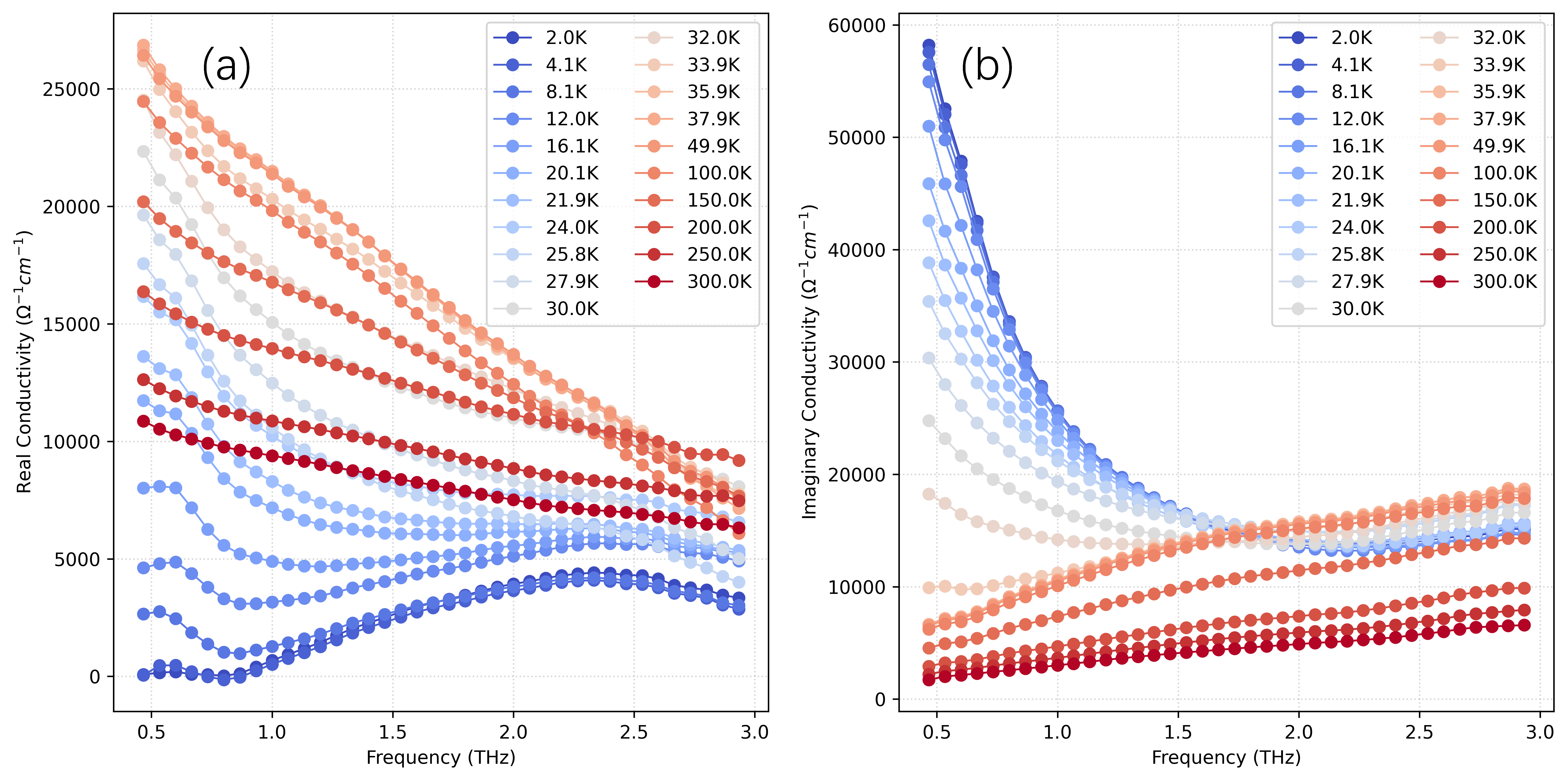}
    \caption{The (a) real and (b) imaginary parts of the optical conductivity of MgB$_2$ at various temperatures.}
    \label{S3}
\end{figure}

Using the THz field generated by a terahertz antenna, we measure the time-domain waveform of the transmitted spectrum for the substrate and sample. Fourier transformation of these signals yields transmission and phase shift in the frequency domain, allowing us to derive the complex optical conductivity. Fig.~S\ref{S2} shows the extracted complex conductivity spectra for MgB$_2$ film. The depletion of $\sigma_{1}(\omega)$ and the $1/\omega$-like behavior of $\sigma_{2}(\omega)$ below $\mathrm{T_c} = 33 \, \text{K}$ indicate gap opening as the sample transitions to a superconducting state. We also conduct temperature-dependent experiments of THz conductivity, as shown in Fig.~S\ref{S3}.

\newpage

\section{Experiment setup}

\begin{figure}[!h]
	\includegraphics[width=16cm]{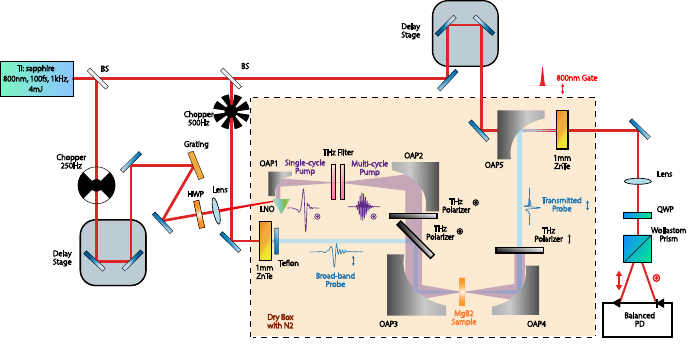}	
    \renewcommand{\figurename}{Fig S}
	\caption{Experimental setup for strong-Field THz pump-Probe spectroscopy: BS - Beam Splitter, OAP - Off-Axis Parabolic Mirror, LNO - LiNbO$_3$ Crystal, HWP - Half-Wave Plate, QWP - Quarter-Wave Plate.}
    \label{Fig:S4}
\end{figure}

The THz pump-probe experiment set-up schematic is shown in Fig.~S\ref{Fig:S4}. A Ti-sapphire laser emitting pulses at 800 nm, with a repetition rate of 1 kHz, pulse duration of 100 fs, and single-pulse energy of 4 mJ, is split into three beams at ratios of 80:18:2 after passing through two beam splitters. These beams are utilized to either generate or act as the THz pump, the THz probe, or the 800 nm gate, respectively. In the pump path, the laser is directed onto a grating after being modulated by a delay stage and a 250 Hz chopper. This process imparts a tilt to the wavefront of the 800 nm pulse, which is then directed onto a lithium niobate (LiNbO$_3$) crystal to generate a strong-field THz pump\cite{yeh_generation_2007}. After passing through three off-axis parabolic mirrors and two polarizers, this pulse is focused onto the sample with vertical electric polarization. In multicycle pump experiments, two THz band-pass filters (THz-BPF) are used to filter out the multicycle THz pump. In contrast, a 1 mm-thick silicon film is employed to block the 800 nm pulse for single-cycle pump experiments. In the probe path, the laser undergoes modulation by a 500 Hz chopper and is then directly incident upon a 1 mm-thick ZnTe crystal, generating weak-field probe THz through optical rectification. After passing through a THz polarizer and an off-axis parabolic (OAP) mirror, the THz probe is focused onto the sample with horizontal electric polarization. With orthogonal electric polarization directions, the THz pump and probe beams traverse the sample and generate a THz nonlinear signal. A polarizer, oriented orthogonal to the pump's polarization, filters out its main component before the beam is focused on the ZnTe sampling crystal. The beam is aligned co-linearly with the 800 nm gate pulse, which is modulated by another delay stage. Ultimately, the THz signal is acquired via electro-optic sampling.

\section{Spectrum of THz pump and THz probe}

\begin{figure}[!h]
    \centering
    \renewcommand{\figurename}{Fig S}

    \begin{subfigure}[b]{0.45\linewidth}
        \centering
        \includegraphics[width=\linewidth]{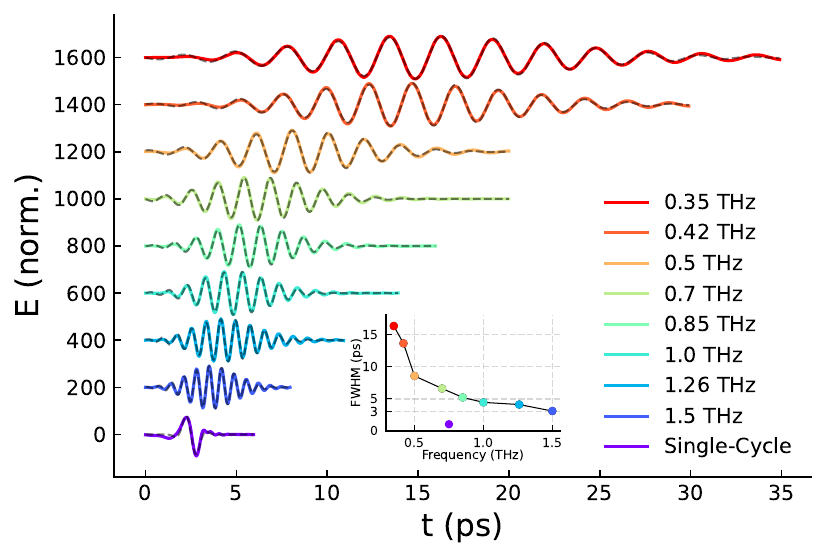}
        \caption{Time domain spectrum}
        \label{fig:THz_pump_time}
    \end{subfigure}
    \begin{subfigure}[b]{0.45\linewidth}
        \centering
        \includegraphics[width=\linewidth]{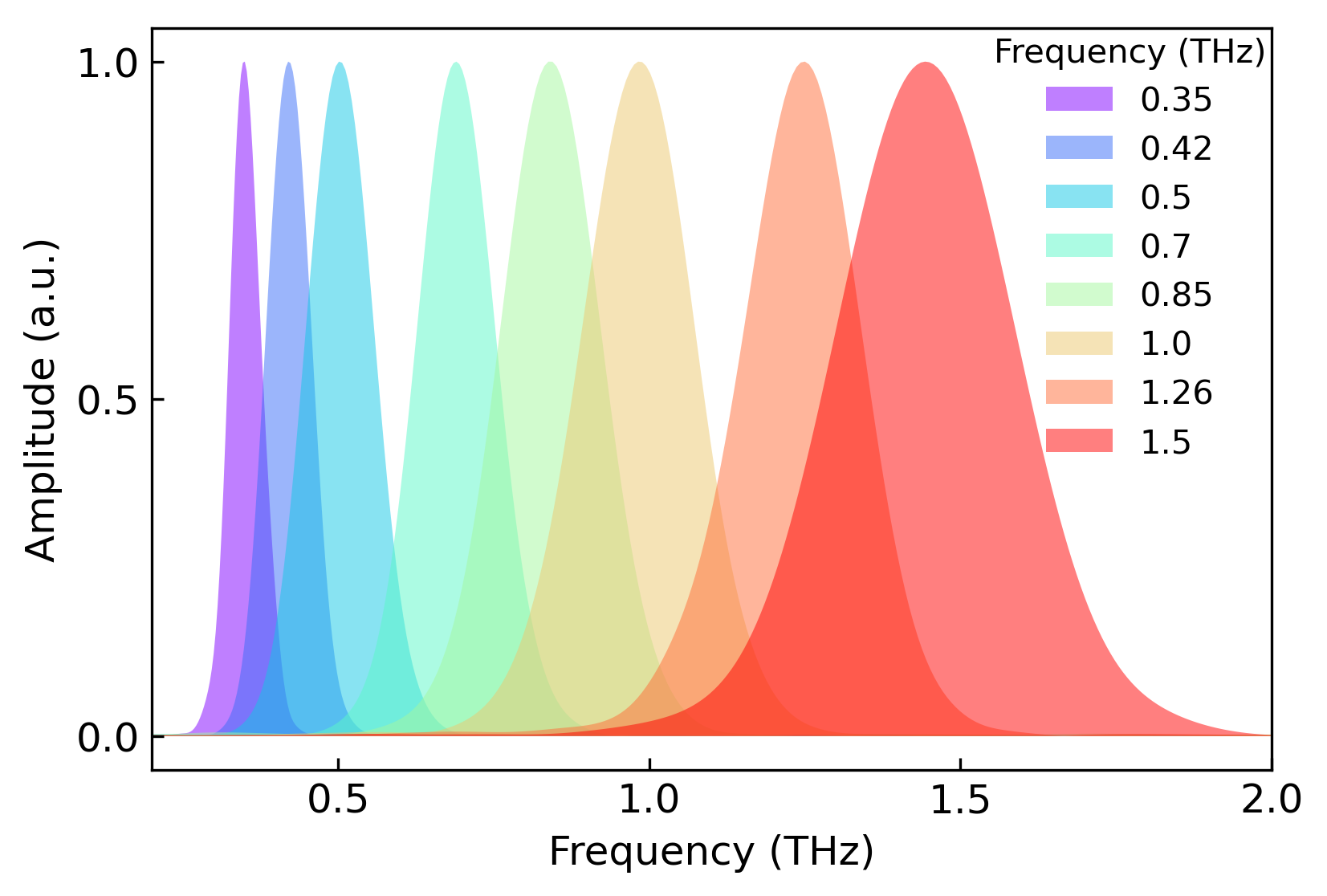}
        \caption{Frequency domain spectrum}
        \label{fig:THz_pump_freq}
    \end{subfigure}
    
    \caption{Spectrum of THz pump.}
    \label{fig:THz_pump}
\end{figure}

\begin{figure}[!h]
    \centering
    \renewcommand{\figurename}{Fig S}

    \begin{subfigure}[b]{0.5\linewidth}
        \centering
        \includegraphics[width=\linewidth]{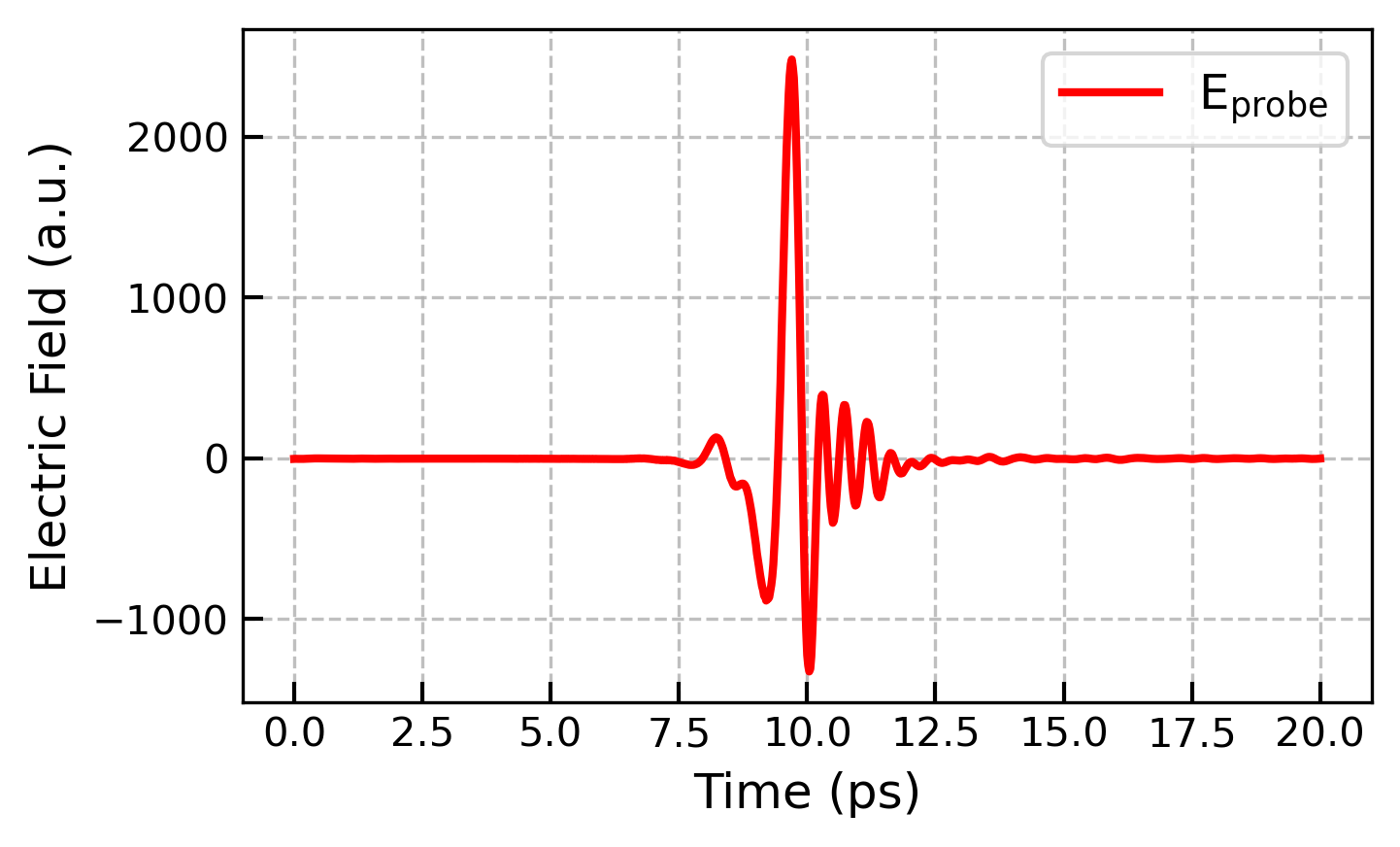}
        \caption{Time domain spectrum}
        \label{fig:THz_probe_time}
    \end{subfigure}
    \begin{subfigure}[b]{0.4\linewidth}
        \centering
        \includegraphics[width=\linewidth]{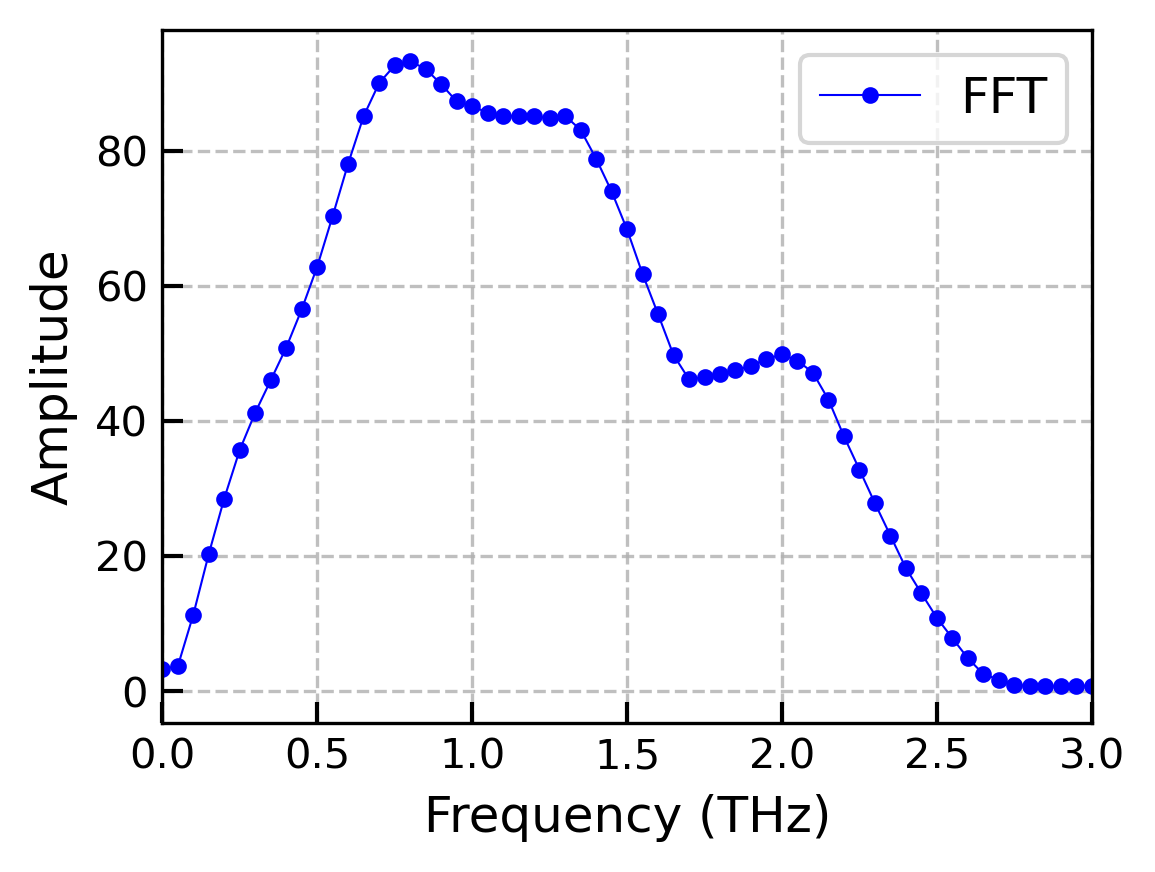}
        \caption{Frequency domain spectrum}
        \label{fig:THz_probe_freq}
    \end{subfigure}
    
    \caption{Spectrum of THz probe generated by ZnTe crystal.}
    \label{fig:THz_probe}
\end{figure}

\section{Logical of Pumpprobe 2D Scan}

\begin{figure}[!h]
    \centering
    \renewcommand{\figurename}{Fig S}
    \includegraphics[width=0.4\linewidth]{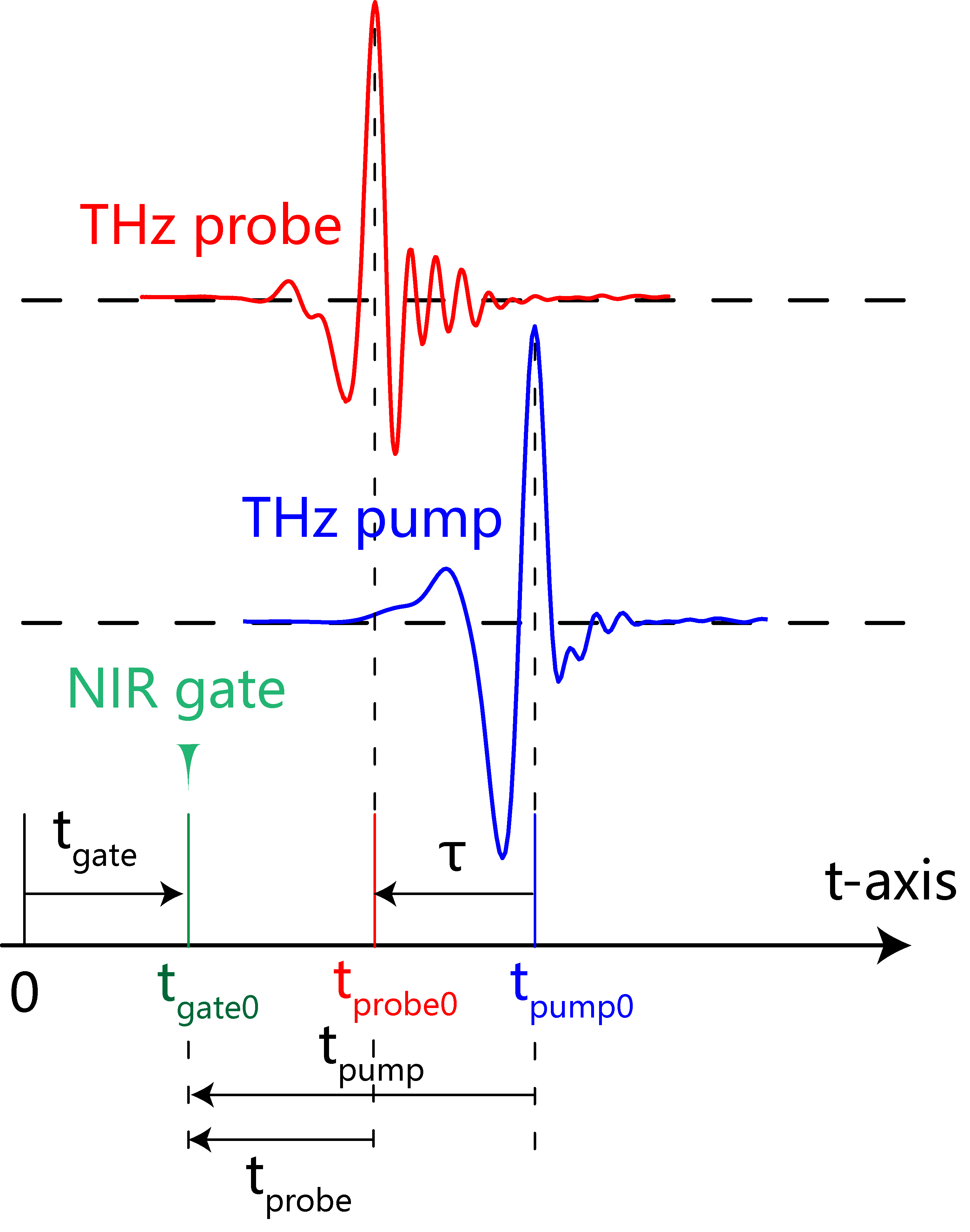}
    \caption{Two Schemes for THz 2D Scan.}
    \label{Fig:S6b}
\end{figure}

The definition of [$\mathrm{t_{pump}}$,$\mathrm{t_{probe}}$] and [$\mathrm{\tau}$,$\mathrm{t_{gate}}$] is shown in Fig.~S\ref{Fig:S6b}.
$\mathrm{t_{pump}}$ is defined as the position of the gate pulse $\mathrm{t_{gate0}}$ relative to the zero-time point of the pump pulse $\mathrm{t_{pump0}}$, whereas $\mathrm{t_{probe}}$ is defined as the position of the gate pulse $\mathrm{t_{gate0}}$ relative to the zero-time point of the probe pulse $\mathrm{t_{probe0}}$. Unlike $\mathrm{\tau}$, which in traditional two-dimensional THz spectroscopy refers to the relative position between the zero-time points of the probe $\mathrm{t_{probe0}}$ and pump pulses $\mathrm{t_{pump0}}$.  The relationship between them is expressed as:

\begin{equation}
\tau = \mathrm{t_{pump}} - \mathrm{t_{probe}}
\end{equation}

\section{Double Modulate Data Acquisition}

\begin{figure}[!h]
	\includegraphics[width=16cm]{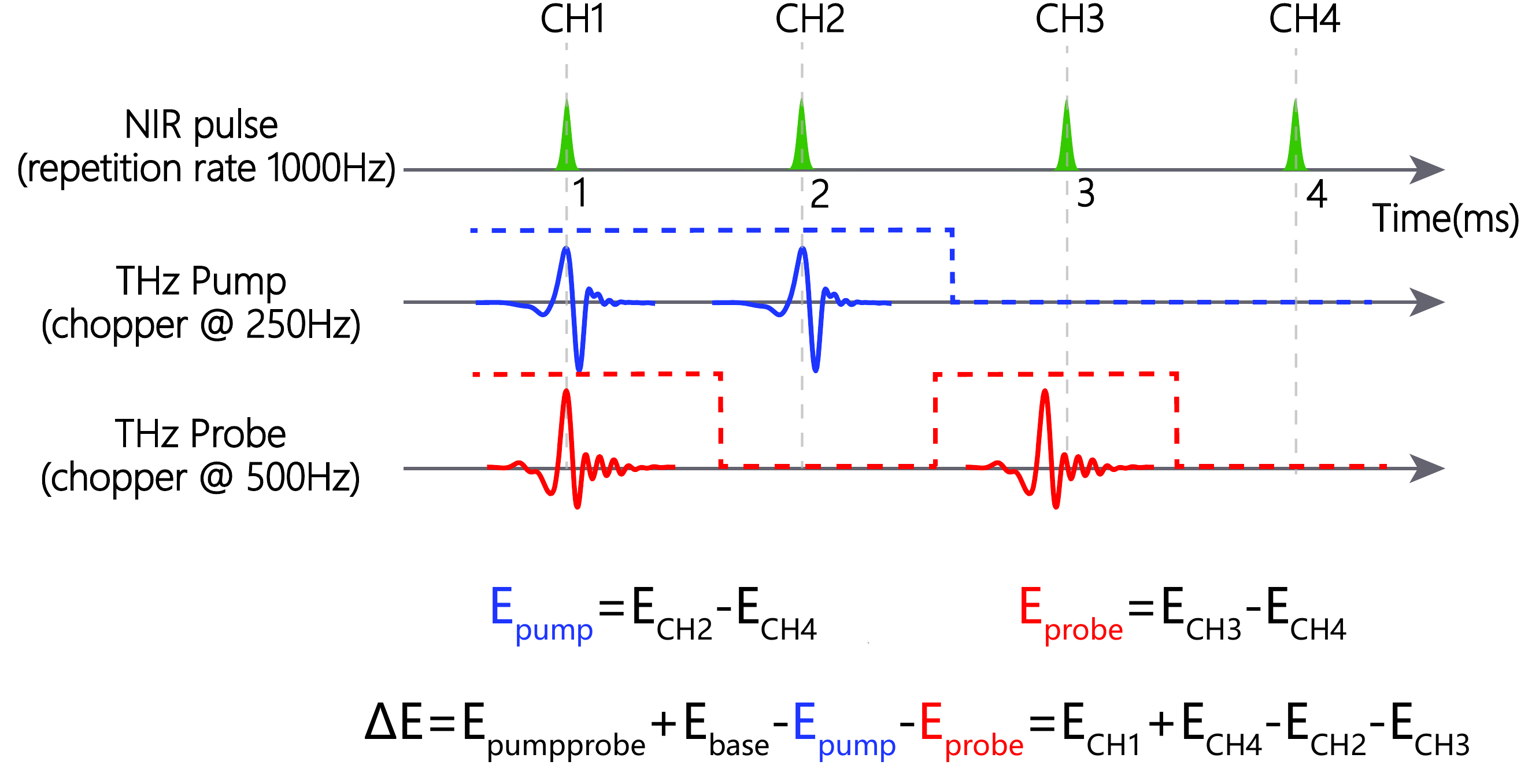}	
    \renewcommand{\figurename}{Fig S}
	\caption{Principle of nonlinear signal acquisition through dual modulation sampling. A differential signal is captured by a data acquisition card from a balanced detector, featuring a sampling period of 4 ms.}
    \label{Fig:S5}
\end{figure}

The schematic of dual modulation sampling is shown in Fig.~S\ref{Fig:S5}.
The laser emits an 800 nm pulse at a repetition rate of 1000 Hz. Modulation occurs at 250 Hz for the pump path and 500 Hz for the probe path, leading to a minimum experimental period of 4 ms. This period accommodates four sampling instances. On the first pulse, the chopper permits both pump and probe beams, enabling the balanced detector to detect $\mathrm{E_{pump}}$, $\mathrm{E_{probe}}$, and $\Delta$E signals. The chopper's blocking results in the second pulse detecting Epump, the third pulse Eprobe, and the fourth pulse capturing the baseline. As a result, the nonlinear signal $\mathrm{\Delta E}$ is calculated by the formula $\mathrm{\Delta E = E_{\text{CH1}} + E_{\text{CH4}} - E_{\text{CH2}} - E_{\text{CH3}}}$. The signal is collected by a data acquisition card and processed with a sampling time of 1 s for each point, where each point corresponds to the fixed positions of two translation stages, allowing for an average of 250 repetitions for each $\mathrm{\Delta E}$ signal.

\section{2$\omega$ Spectral Weight Fluence Dependence at various frequencies}

\begin{figure}[!h]
	\includegraphics[width=0.8\linewidth]{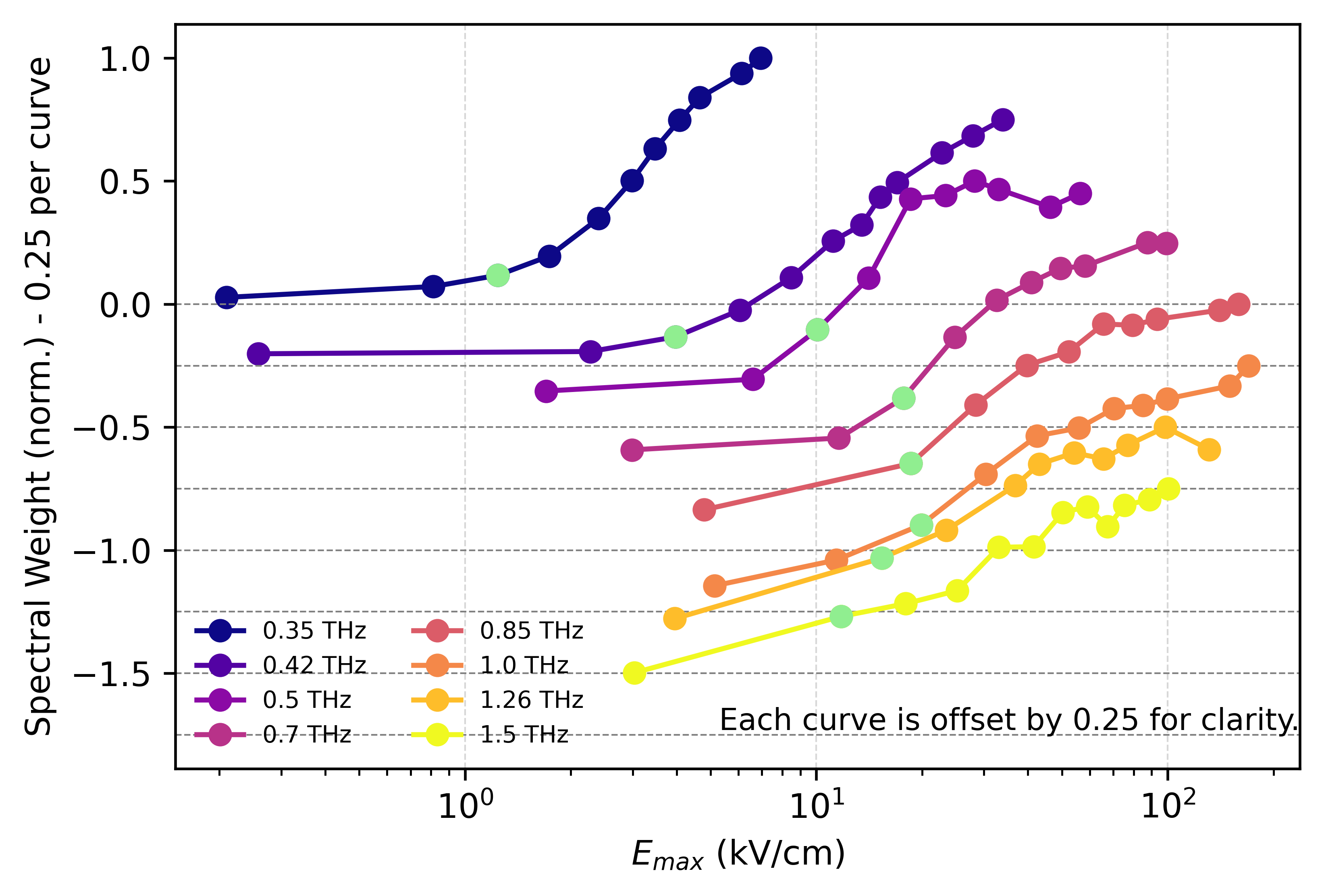}	
    \renewcommand{\figurename}{Fig S}
	\caption{2$\omega$ Spectral Weight Fluence Dependence at different pump frequencies. Green points show the selected fluency in calculating and plotting Main-text Fig 2B at corresponding pump frequencies.}
    \label{Fig:S14}
\end{figure}

We conduct multicycle pump-probe fluence dependence experiments for all driving frequencies. Based on the experimental results, we calculated the fluence dependence of the 2$\omega$ spectral weight for each frequency, which is presented in Fig.~S\ref{Fig:S14}. The green dots in the figure represent the fluence values corresponding to each frequency used in plotting Fig 2B in the main text.

\section{Calculation of Transmission Coefficients from Complex Refractive Indices and Considering Screening Effect}

In this section, we discuss the calculation of transmission coefficients for light transitioning from dry air into the sample across various terahertz frequencies. The transmission coefficient, $ \mathrm{Tr} $, is calculated using the formula:

\begin{equation}
\mathrm{Tr} = \left| \frac{2}{1 + N} \right|
\end{equation}

\begin{figure}[htbp]
\centering
\renewcommand{\figurename}{Fig S}
\begin{subfigure}[b]{0.4\linewidth}
    \centering
    \includegraphics[width=\linewidth]{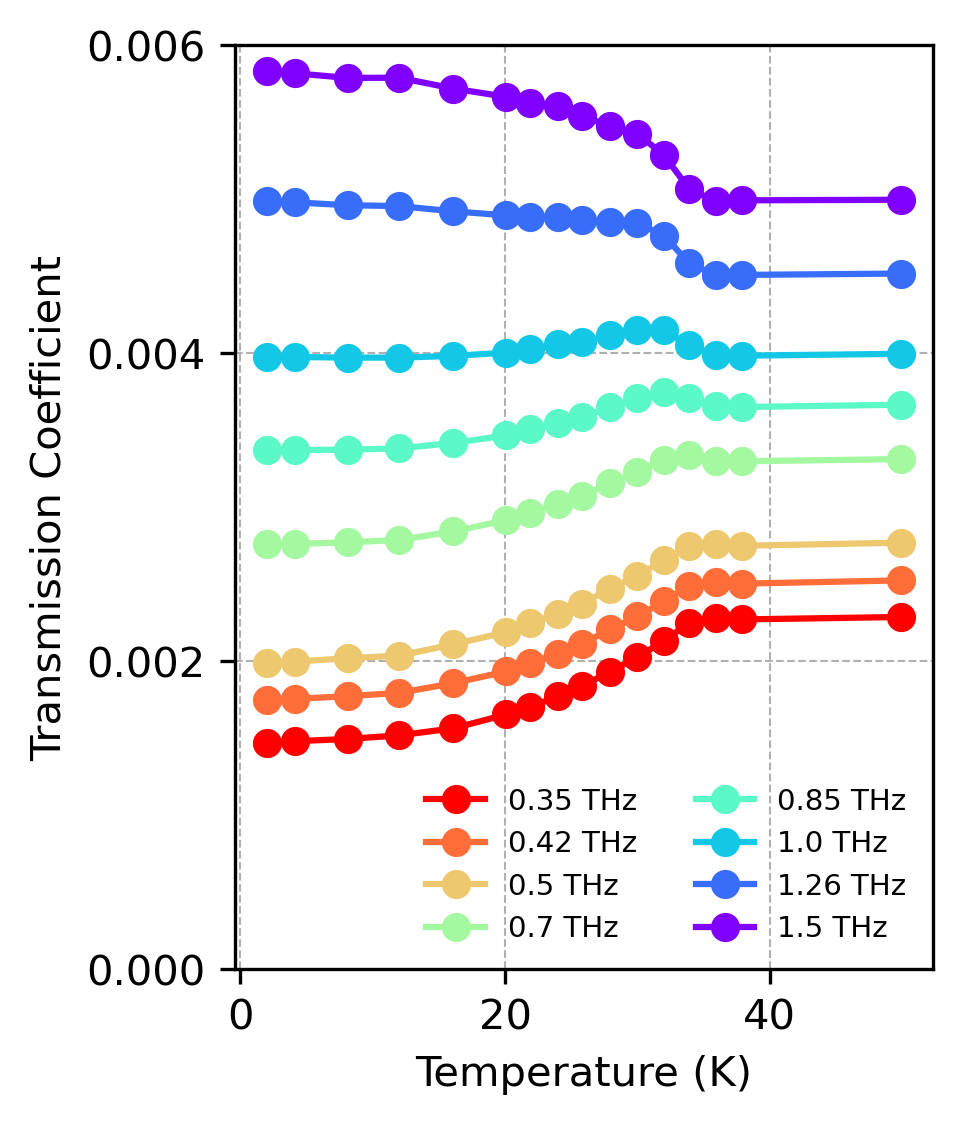}
    \caption{Transmission coefficients as a function of temperature.}
    \label{fig:S11a}
\end{subfigure}
\begin{subfigure}[b]{0.4\linewidth}
    \centering
    \includegraphics[width=\linewidth]{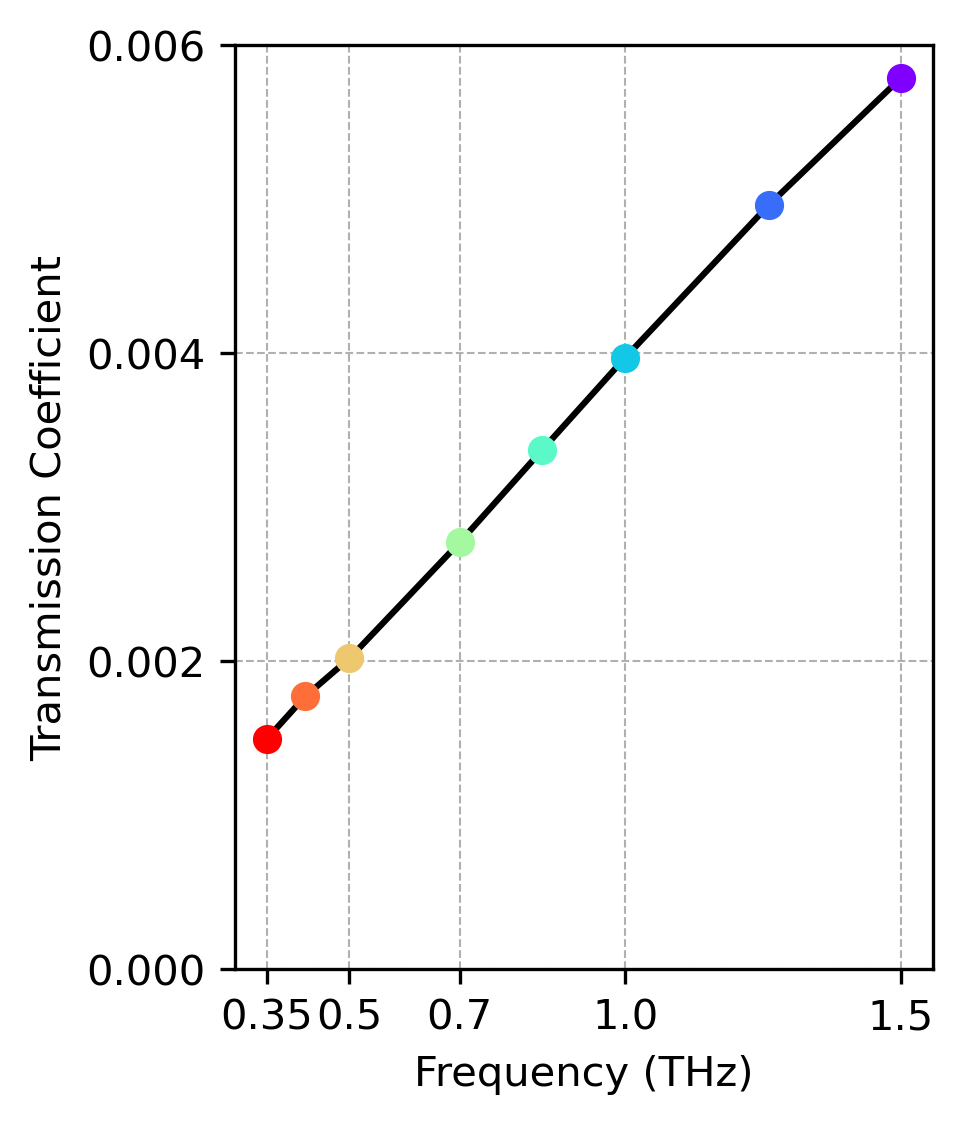}
    \caption{Transmission coefficients as a function of frequency at 8K}
    \label{fig:S11b}
\end{subfigure}
\caption{Transmission coefficient for normal incidence, calculated for the air-to-sample interface using THz-TDS measurements.}
\label{fig:S11}
\end{figure}

Where $ N $ is the complex refractive index, defined as $ N = n + i \cdot k $, where $ n $ is the refractive index and $ k $ is the extinction coefficient. The refractive index and extinction coefficient are determined based on terahertz time-domain spectroscopy (THz-TDS) measurements. The graph shown in Fig.~S\ref{fig:S11a} illustrates how the transmission coefficients vary with temperature for different frequencies.

To account for the screening effect of the pump, the normalized nonlinear signals are calculated by dividing the measured signals by the appropriate powers of the pump transmission $\mathrm{Tr}(\omega_{pump})$, depending on the number of pump interactions involved. The expressions are as follows:

\begin{equation}
\begin{aligned}
    \mathrm{\Delta E}_{1\omega}^{\text{norm}}(T) &= \frac{\mathrm{\Delta E}_{1\omega}(T)}{\mathrm{Tr}(\omega_{pump},T)\mathrm{Tr}^2(\omega_{probe},T)}, \\
    \mathrm{\Delta E}_{2\omega}^{\text{norm}}(T) &= \frac{\mathrm{\Delta E}_{2\omega}(T)}{\mathrm{Tr}^2(\omega_{pump},T)\mathrm{Tr}(\omega_{probe},T)}.
\end{aligned}
\end{equation}

Here, $\mathrm{\Delta E}_{1\omega}^{\text{norm}}$ corresponds to the nonlinear susceptibility term $\chi^{(3)}(\omega_{pump},\omega_{probe},-\omega_{probe})$, involving a single interaction with the pump, so it is divided by $\mathrm{Tr}(\omega_{pump})\mathrm{Tr}^2(\omega_{probe})$. Similarly, $\mathrm{\Delta E}_{2\omega}^{\text{norm}}$ corresponds to the susceptibility term $\chi^{(3)}(\omega_{pump},\omega_{pump},\omega_{probe})$, which involves two pump interactions, so it is divided by $\mathrm{Tr}^2(\omega_{pump})\mathrm{Tr}(\omega_{probe})$. Note that, referring to the $\mathrm{E_{probe}}$ spectrum, whose central frequency is at 1.0 THz, we simplify the consideration of $ \mathrm{Tr}(\omega_{{probe}}) $ by using $ \omega_{{probe}} = 1.0 $ THz.

\begin{figure}[!h]
    \includegraphics[width=0.6\textwidth]{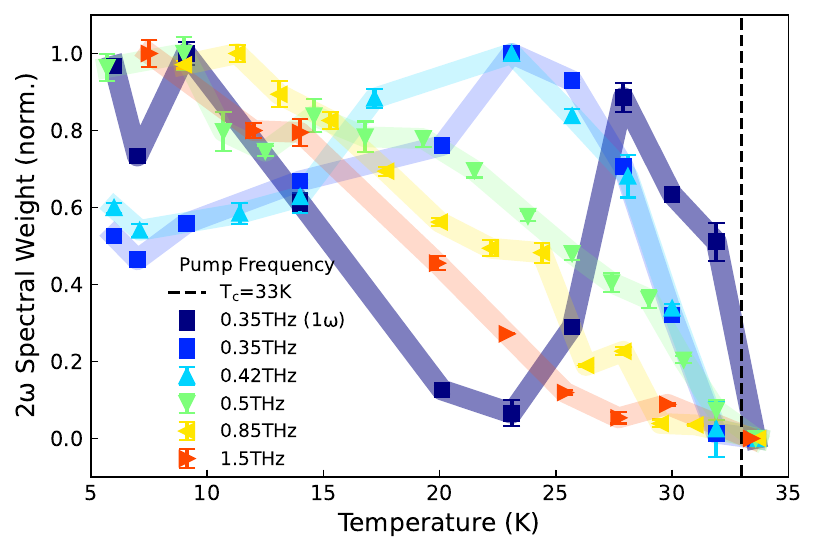}    
    \renewcommand{\figurename}{Fig S}
    \caption{2$\omega$ Spectra Weight versus Temperature at Various Pump Frequency without Considering Screening Effect.
}
    \label{Fig:S12}
\end{figure}

It can be observed that the resonance behavior of 2$\omega$ oscillation at 0.35 and 0.42 THz is apparent before considering the screening effect, as shown in Fig.~S\ref{Fig:S12}. 

\section{Time-Domain and Frequency-Domain Nonlinear Signals at Different Temperatures for Various Driving Frequencies}

\begin{figure}[!h]
    \centering
    \renewcommand{\figurename}{Fig S}
    
    \begin{subfigure}[b]{0.4\linewidth}
        \centering
        \caption{\textbf{0.35 THz}}
        \includegraphics[width=\linewidth]{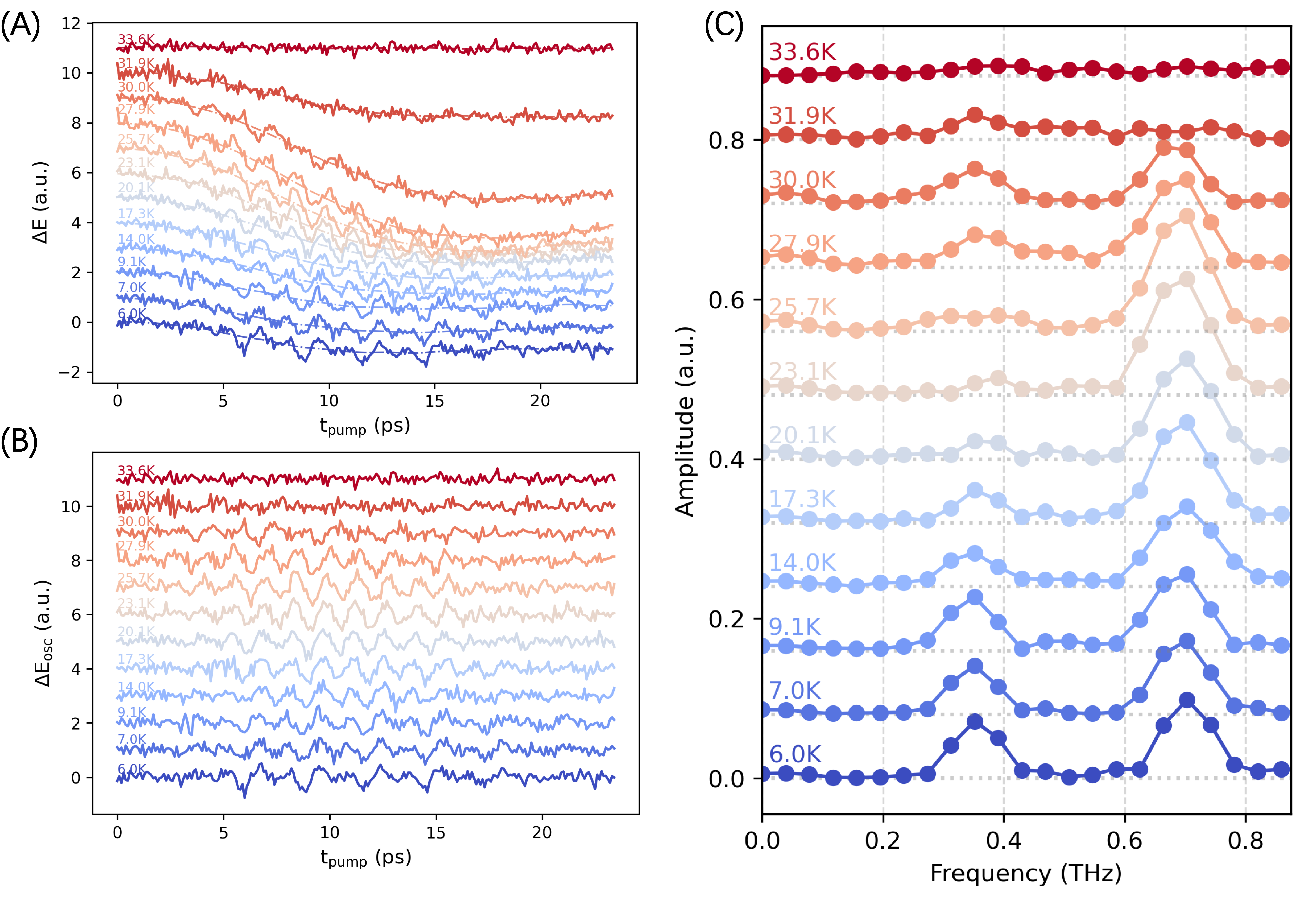}
        \label{Fig:0.35}
    \end{subfigure}%
    \begin{subfigure}[b]{0.4\linewidth}
        \centering
        \caption{\textbf{0.42 THz}}
        \includegraphics[width=\linewidth]{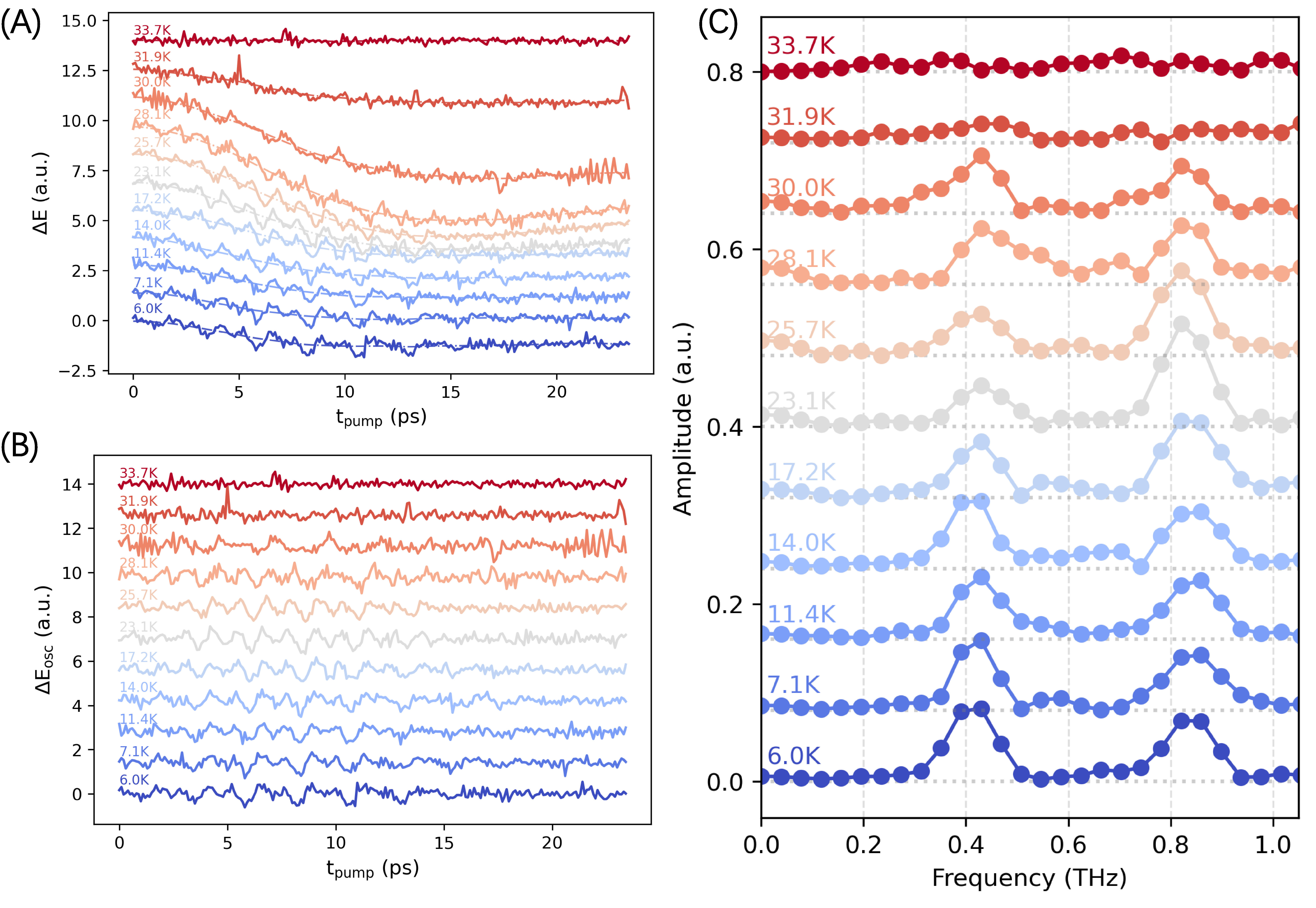}
        \label{Fig:0.42}
    \end{subfigure}
    
    \begin{subfigure}[b]{0.4\linewidth}
        \centering
        \caption{\textbf{0.5 THz}}
        \includegraphics[width=\linewidth]{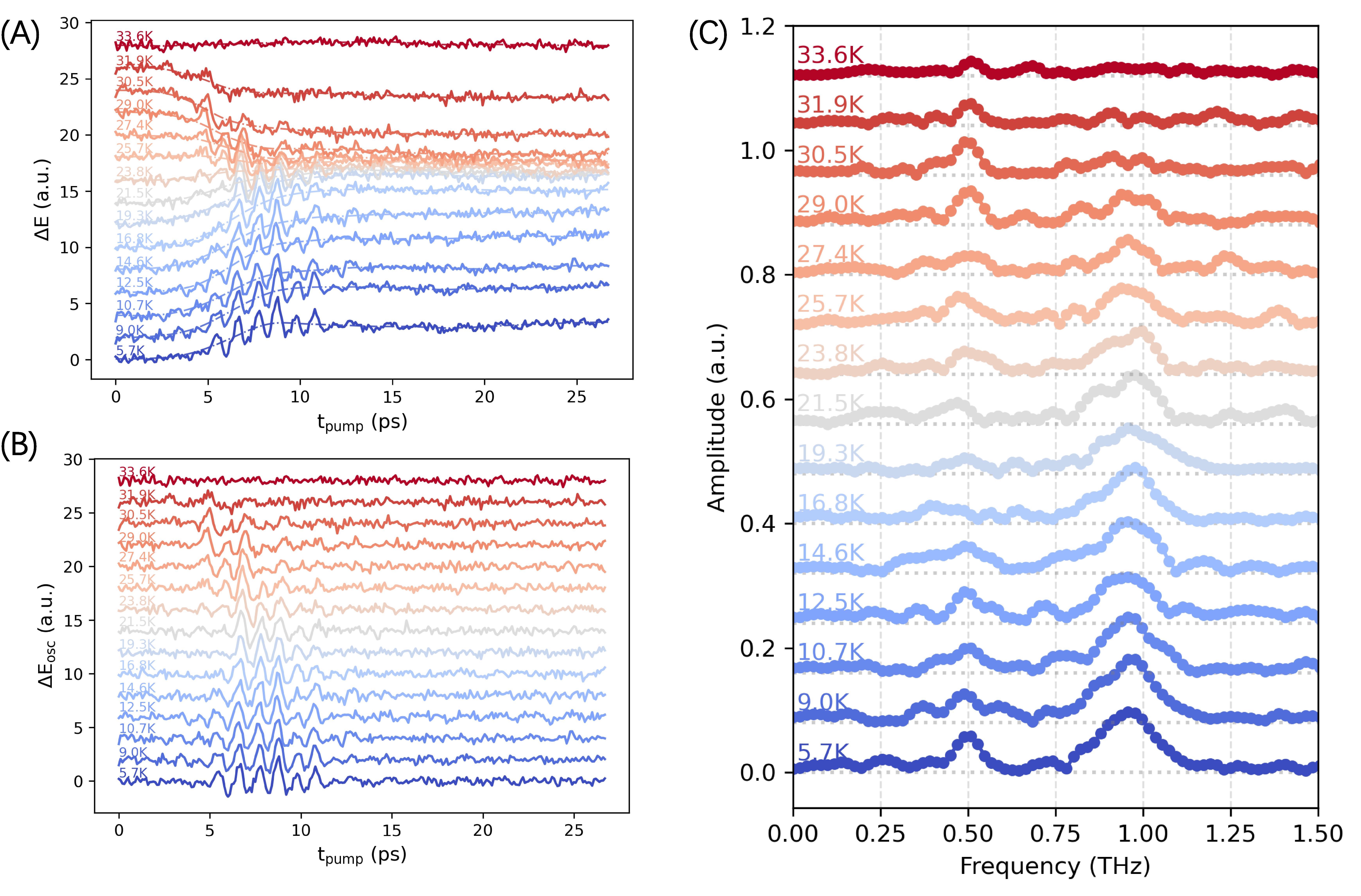}
        \label{Fig:0.5}
    \end{subfigure}%
    \begin{subfigure}[b]{0.4\linewidth}
        \centering
        \caption{\textbf{0.85 THz}}
        \includegraphics[width=\linewidth]{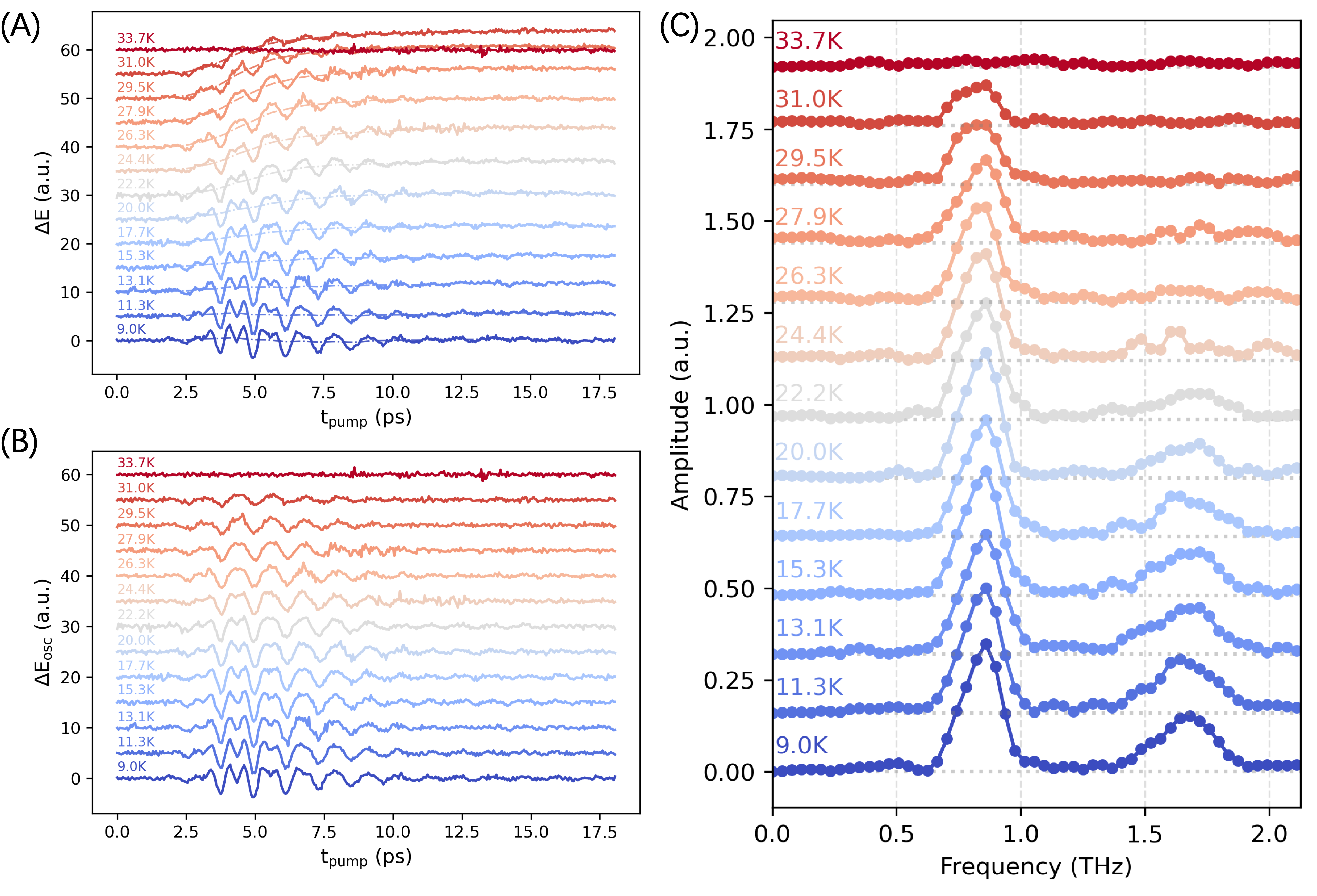}
        \label{Fig:0.85}
    \end{subfigure}%

    \begin{subfigure}[b]{0.4\linewidth}
        \centering
        \caption{\textbf{1.5 THz}}
        \includegraphics[width=\linewidth]{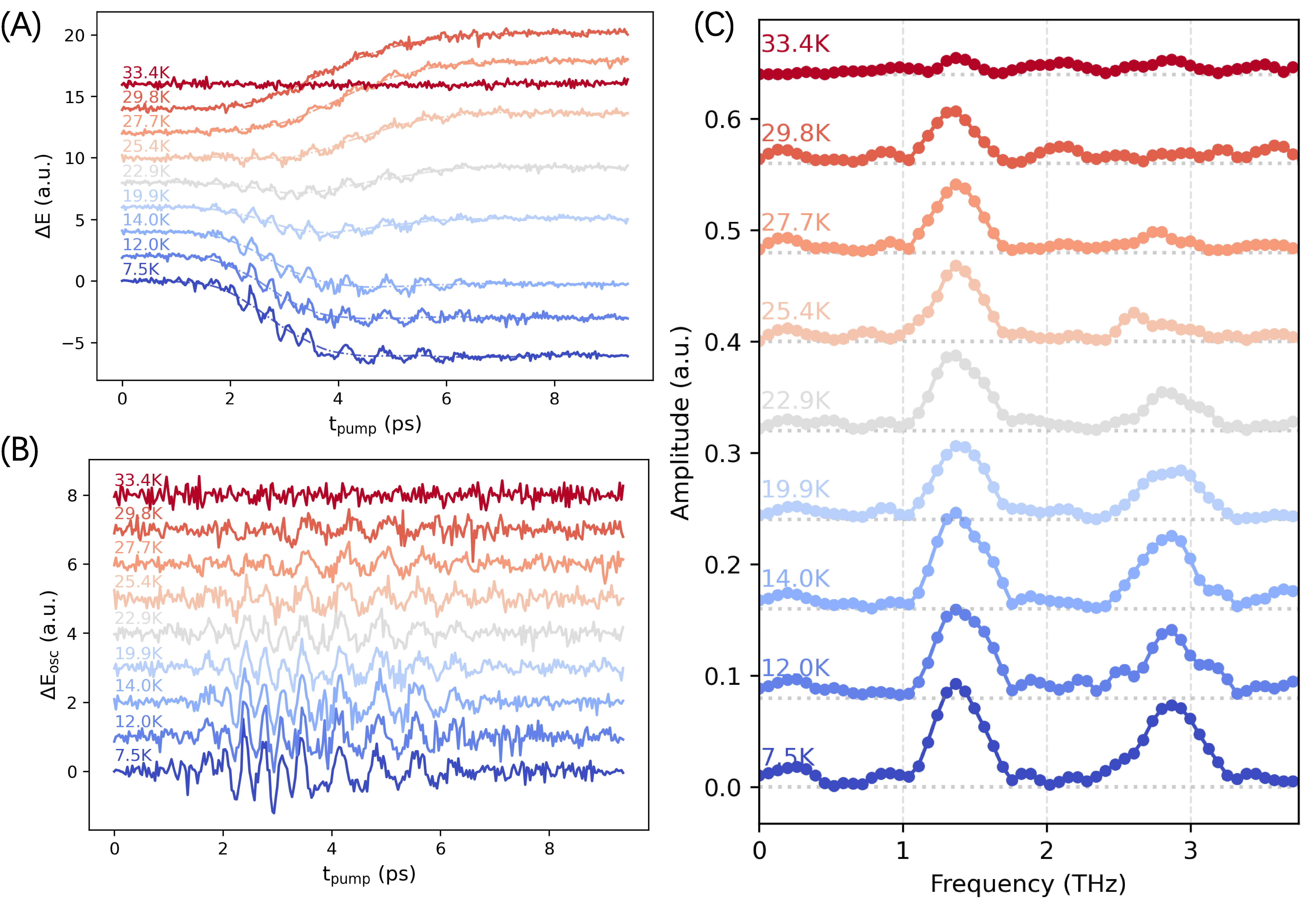}
        \label{Fig:1.5}
    \end{subfigure}
    
    \caption{Time-domain and frequency-domain analysis at different frequencies (0.35 THz, 0.42 THz, 0.5 THz, 0.85 THz, and 1.5 THz). The subfigures (a), (b), (c), (d), and (e) correspond to the data at specific driving frequencies. Each subfigure contains three panels: A: original time-domain spectra at various temperatures, B: oscillatory component after background subtraction, and C: FFT of the oscillatory component.}

    \label{Fig:combined}
\end{figure}

We present the original time-domain and frequency-domain nonlinear signals measured at various temperatures and driving frequencies, as shown in Fig.~S\ref{Fig:combined}. These signals are used to calculate $1\omega$ and $2\omega$ spectral weight, which are shown in Fig.~2F of the main text.

\section{Calculation of Pump Maximum Electric Field}

The peak electric field strength of the driving pump pulses is determined using the formula:

\begin{equation}
    W = \epsilon_0 c A \int_{-\infty}^{\infty} E(t)^2 \, dt = \epsilon_0 c A E_{\text{max}}^2 \int_{-\infty}^{\infty} \left( \frac{E(t)}{E_{\text{max}}} \right)^2 \, dt
\end{equation}

In this equation, $ W $ denotes the single-pulse energy of the terahertz pulse at the focal point, while $ A $ represents a normalization coefficient with dimensions of area. After normalizing with the peak electric field $ E_{\text{max}} $, the formula for the peak electric field strength is expressed as:

\begin{equation}
    E_{\text{max}} = \sqrt{\frac{W}{\epsilon_0 c A \int_{-\infty}^{\infty} \left( \frac{E(t)}{E_{\text{max}}} \right)^2 \, dt}}
\end{equation}

The parameter $ A $ is calculated based on the full width at half maximum (FWHM) of the terahertz spot along the horizontal axis, denoted as $\mathbf{a}$, and the vertical axis, denoted as $\mathbf{b}$, from a two-dimensional Gaussian distribution. The expression for $ A $ is given as:

\begin{equation}
    A = \frac{\pi a b}{4 \ln 2}
\end{equation}

\begin{figure}[!h]
    \includegraphics[width=16cm]{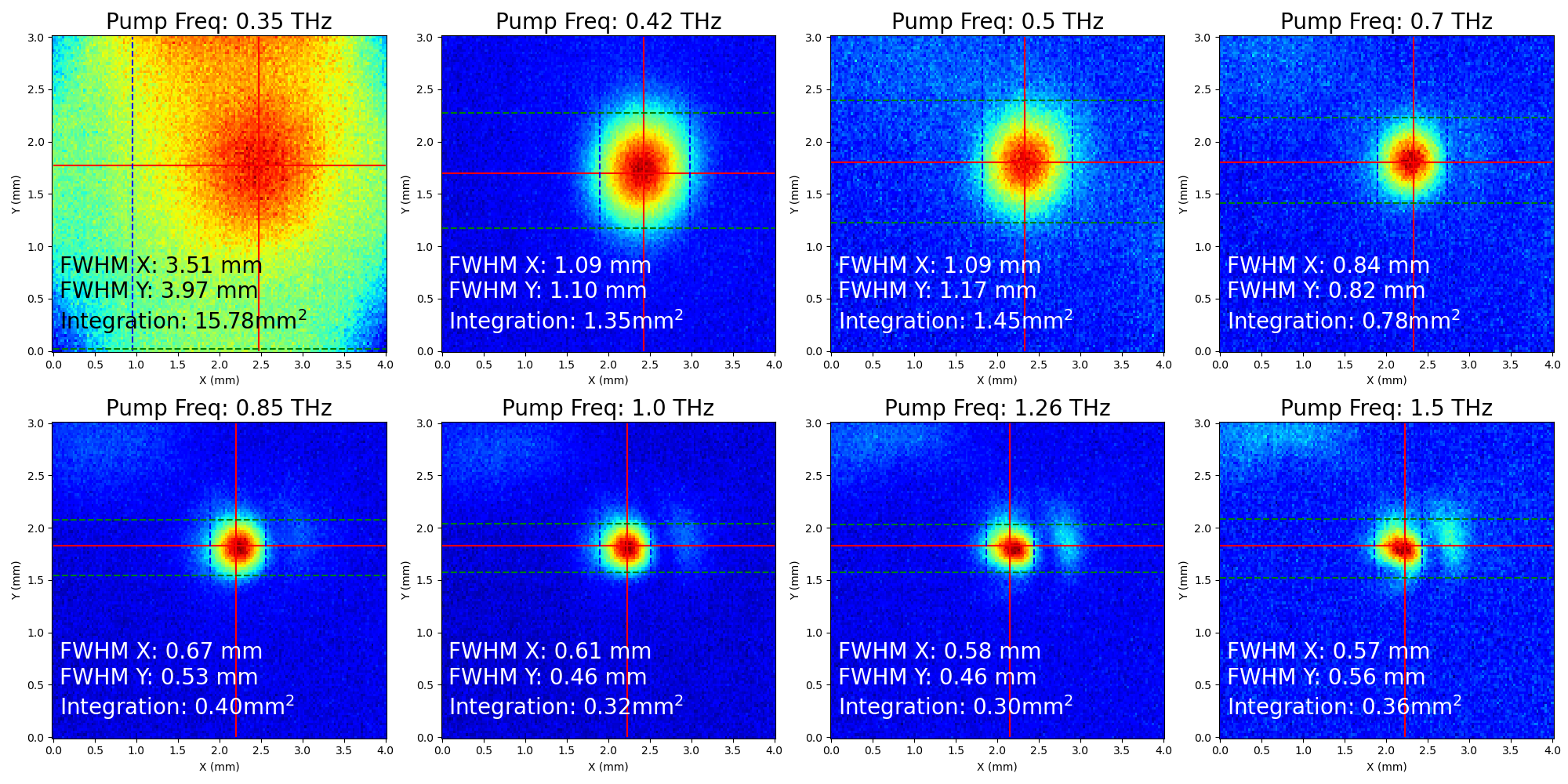}	
    \renewcommand{\figurename}{Fig S}
    \caption{
    The terahertz camera captures the distribution of the multi-cycle pump's spot on the sample at various frequencies. The full width at half maximum (FWHM) values along the horizontal and vertical directions, obtained via two-dimensional Gaussian fitting, are annotated, along with the integrated value A.}
    \label{Fig:S13}
\end{figure}

The distribution of the multi-cycle pump's spot is illustrated in Fig.~S\ref{Fig:S13}, which shows the spot distribution captured by the terahertz camera at the focal plane. The FWHM values and the area integral of the probability amplitude are annotated in the lower-left corner of the figure.

The FWHM values along the $ x $- and $ y $-directions are obtained through two-dimensional Gaussian fitting. The energy of the time-domain electric field waveform $ E(t) $ at the focal plane, measured via electro-optic sampling, is quantified by the integral $ \int_{-\infty}^{\infty} E(t)^2 \, dt $, normalized by the square of the peak electric field $ E_{\text{max}}^2 $.

Additionally, the ratio of the electric field at a specific polarizer angle to the peak electric field $ E_{\text{max}} $ is calculated. This allows us to determine the peak electric field strength inside the sample, denoted as $ E_{\text{max}}^{\text{in}} $:

\begin{equation}
    E_{\text{max}}^{\text{in}} = E_{\text{max}} \times \cos^2\theta \times \text{Tr} 
\end{equation}

In this equation, $ \text{Tr} $ represents the normal-incidence transmission coefficient at the air-to-sample interface, determined via THz-TDS measurements, while $ \cos^2\theta $ indicates the relative intensity, with $ \theta $ being the angle between two polarizers.

The calculated parameters, including the peak electric field strengths at various terahertz frequencies, are presented in Table.~\ref{tab:s2} (where $W$ in our experiment is measured using a thermal THz power meter).

\begin{table}[!h]
    \centering
    \renewcommand{\thetable}{S \arabic{table}}
    \begin{tabular}{rrrrrrrr}
    \toprule
    Freq (THz) &   $\int{E^2dt}$ ($E^2_{\text{max}}$ s) &   A (mm$^2$) & W (nJ) & $E_{\text{max}}$ (kV/cm) & $\cos^2\theta$ & Tr & $E^{\text{trans}}_{\text{max}}$ (kV/cm)\\
    \midrule
    0.35 &  6.28 &  15.8 & 144 & 7.0   & 0.18 & 1.5e-3 & 1.9e-3\\
    0.42 &  5.21 &  1.35 & 244 & 33.9  & 0.12 & 1.7e-3 & 6.9e-3\\
    0.50 &  3.21 &  1.45 & 445 & 56.5  & 0.18 & 2.0e-3 & 2.0e-2\\
    0.70 &  2.53 &  0.78 & 580 & 99.0  & 0.18 & 2.7e-3 & 4.9e-2\\
    0.85 &  1.98 &  0.40 & 605 & 158.9 & 0.12 & 3.4e-3 & 6.3e-2\\
    1.00 &  1.76 &  0.32 & 490 & 170.0 & 0.12 & 4.0e-3 & 7.9e-2\\
    1.26 &  1.55 &  0.30 & 240 & 131.3 & 0.12 & 5.0e-3 & 7.7e-2\\
    1.50 &  1.18 &  0.36 & 130 & 100.7 & 0.12 & 5.8e-3 & 6.9e-2\\
    \bottomrule
    \end{tabular}
    \caption{Parameters and peak electric field strengths at the focal position for the multi-cycle terahertz pump across different frequencies.}
    \label{tab:s2}
\end{table}

\section{Calculation of Nonlinear Coefficient $\mathrm{\chi^{(3)}}$}

After determining the peak electric field of the driving pump pulse inside the sample, the next step is to calculate the actual peak electric field of the 2$\omega$ oscillations. This value is obtained through electro-optic sampling and is corrected for dispersion effects introduced during the sampling process. 

Since our goal is to compare the nonlinear coefficients within the frequency range of 0.35 THz to 1.5 THz, it is sufficient to determine the normalized peak electric field of the corrected 2$\omega$ oscillations. To achieve this, we first derive the formula that converts the differential signal $\Delta I$ obtained from electro-optic sampling into the terahertz electric field $E_\text{THz}$, as shown below\cite{casalbuoni_numerical_2008}:

\begin{equation}
E_{\mathrm{THz}} = \frac{\lambda}{2\pi n^3 r_{41}(\omega) {\mathrm{t_{ZnTe}}(\omega)} L G(\omega)} 
\sin^{-1}\left(\frac{\Delta I}{I}\right)
\end{equation}

For small signal conditions where $\Delta I \ll I$, the relationship simplifies to:

\begin{equation}
E_{\mathrm{THz}} \propto \frac{\Delta I}{r_{41}(\omega) {\mathrm{t_{ZnTe}}(\omega)} G(\omega)} = \eta(\omega)\Delta I
\end{equation}

Here, the parameter $\eta(\omega)$ represents the dispersion factor in the terahertz range. To determine $\eta(\omega)$, we use the mismatch correction response $ G(\omega) $, the electro-optic coefficient $ r_{41} $, and the terahertz transmission coefficient ${\mathrm{t_{ZnTe}}(\omega)}$, with values obtained from referenced data \cite{casalbuoni_numerical_2008}. 

By calculating the normalized dispersion factor $\eta(\omega)$ during the sampling process, we can combine it with the measured amplitude of the 2$\omega$ oscillations from electro-optic sampling. This allows us to derive the normalized amplitude of the 2$\omega$ terahertz electric field at each driving pulse frequency. The results are summarized in the Table.~\ref{tab:s3}.

\begin{table}[!h]
\centering
\renewcommand{\thetable}{S \arabic{table}}
\begin{tabular}{rrrrrrr}
\toprule
 Frequency (THz) &   ${\mathrm{t_{ZnTe}}(\omega)}$ &  Response &  r41 (m/V) &  Normalized $\eta(\omega)$  & EO Sampling ${E_{2\omega}}$ (a.u.) & Normalized ${E_{2\omega}}$\\
\midrule
            0.70 &                       0.48 &   1.00 & 3.95 e-12 & 0.67 &  0.44 &  0.42 \\
            0.84 &                       0.48 &   1.00 & 3.94 e-12 & 0.67 &  0.41 &  0.40 \\
            1.00 &                       0.48 &   0.98 & 3.94 e-12 & 0.68 &  0.59 &  0.59 \\
            1.40 &                       0.48 &   0.86 & 3.93 e-12 & 0.78 &  0.57 &  0.64 \\
            1.70 &                       0.48 &   0.76 & 3.92 e-12 & 0.88 &  0.78 &  1.00 \\
            2.00 &                       0.47 &   0.69 & 3.90 e-12 & 1.00 &  0.64 &  0.91 \\
            2.50 &                       0.47 &   0.76 & 3.87 e-12 & 0.91 &  0.55 &  0.73 \\
            3.00 &                       0.46 &   0.81 & 3.81 e-12 & 0.89 &  0.46 &  0.59 \\
\bottomrule
\end{tabular}
\caption{Parameters related to the electro-optic sampling process using a ZnTe crystal, the maximum amplitude of the $2\omega$ oscillations obtained from electro-optic sampling (under the flux and temperature conditions used for plotting Fig.~3B in the main text), and the normalized maximum amplitude of the $2\omega$ oscillations after accounting for the dispersion of the terahertz band in the electro-optic sampling system\cite{casalbuoni_numerical_2008}.
}
\label{tab:s3}
\end{table}

Then, combining with the peak electric field strength  $E_{max}$ of the pump pulse estimated in the previous section, we use the formula:
\begin{equation}
    \chi^{(3)}  \propto \frac{E_{2\omega}}{E_{pump}^2E_{probe}}  \propto \frac{E_{2\omega}}{E_{pump}^2} 
\end{equation}
to calculate the nonlinear coefficients at various driving pulse frequencies, as shown in Table.~\ref{tab:s4} and Fig.~3B in the main text.

\begin{table}[!h]
\centering
\renewcommand{\thetable}{S \arabic{table}}
\begin{tabular}{cccc}
\hline
Frequency (THz) & ${E_{2\omega}}$ (a.u.) & ${E^{trans}_{max}}$ (kV/cm) & $\chi^{(3)}$ \\
\hline
0.35 & 0.42 & 1.9e-3 & 1.0 \\
0.42 & 0.40 & 6.9e-3 & 6.8e-2 \\
0.50 & 0.59 & 2.0e-2 & 1.2e-2 \\
0.70 & 0.64 & 4.9e-2 & 2.2e-3 \\
0.85 & 1.00 & 6.3e-2 & 2.1e-3 \\
1.00 & 0.91 & 7.9e-2 & 1.2e-3 \\
1.26 & 0.73 & 7.7e-2 & 1.0e-3 \\
1.50 & 0.59 & 6.9e-2 & 1.0e-3 \\
\hline
\end{tabular}
\caption{Normalized maximum amplitude of $2\omega$ oscillations at various driving pulse frequencies, peak electric field strength at the focus of the driving pulse, and the calculated normalized nonlinear coefficient $\chi^{(3)}$ (as depicted in Fig.~3B of the main text).
}
\label{tab:s4}
\end{table}

\section{Calculation of Leggett Mode Frequency}

The Leggett mode frequency can be calculated using the following formula, based on the intra-band and inter-band pairing potentials, as well as the density of states and energy gaps \cite{blumberg_observation_2007}:

\begin{equation}
\omega_L^2 = \frac{N_\sigma + N_\pi}{N_\sigma N_\pi} \cdot \frac{4V_{\sigma\pi}\Delta_\sigma(T)\Delta_\pi(T)}{\text{det}V}.
\end{equation}

Here, $\mathrm{ V_{\sigma\sigma} }$, $\mathrm{ V_{\pi\pi}} $, and $\mathrm{V_{\sigma\pi}}$ represent the intra-band and inter-band pairing potentials, $\mathrm{N_\sigma}$ and $\mathrm{ N_\pi }$ are the corresponding densities of states, and $\mathrm{\Delta_\sigma(T)}$ and $\mathrm{\Delta_\pi(T)}$ are the temperature-dependent energy gaps.

\begin{table}[!h]
\centering
    \renewcommand{\thetable}{S \arabic{table}}
\begin{tabular}{|c|c|c|c|}
\hline
\textbf{Reference} & $V_{\sigma\sigma} (\text{Ry})$ & $V_{\pi\pi} (\text{Ry})$ & $V_{\sigma\pi} (\text{Ry})$ \\
\hline
Liu et al.\cite{liuEliashbergSuperconductivityMgB2001a} & 0.47 & 0.1 & 0.08 \\
Choi et al.\cite{choiOriginAnomalousSuperconducting2002} & 0.38 & 0.076 & 0.054 \\
Golubov et al.\cite{golubovSpecificHeatMgB22002} & 0.5 & 0.16 & 0.077 \\
\hline
\end{tabular}
\caption{Intra-band ($V_{\sigma\sigma}$, $V_{\pi\pi}$) and inter-band ($V_{\sigma\pi}$) pairing potentials in Ry units from different references\cite{blumberg_observation_2007}.}
\label{tab:s1}
\end{table}

The parameters used in the calculation are as follows: the densities of states are $\mathrm{N_\sigma = 2.04}$ and $\mathrm{N_\pi = 2.78}$ Ry$^{-1}$ spin$^{-1}$ cell$^{-1}$\cite{blumberg_observation_2007}, while the energy gaps are $\mathrm{\Delta_\pi = 0.44 \, \text{THz}}$ and $\mathrm{\Delta_\sigma = 1.32 \, \text{THz}}$, with $\mathrm{\Delta_\sigma}$ assumed to be three times larger than $\mathrm{\Delta_\pi}$. Using these parameters and the three sets of data (as shown in Table.~\ref{tab:s1}), the frequency of the Leggett mode is calculated to be $\mathrm{1.81 \pm 0.27 \, \text{THz}}$, representing its zero-temperature value.

\begin{figure}[!h]
    \centering
    \includegraphics[width=0.5\linewidth]{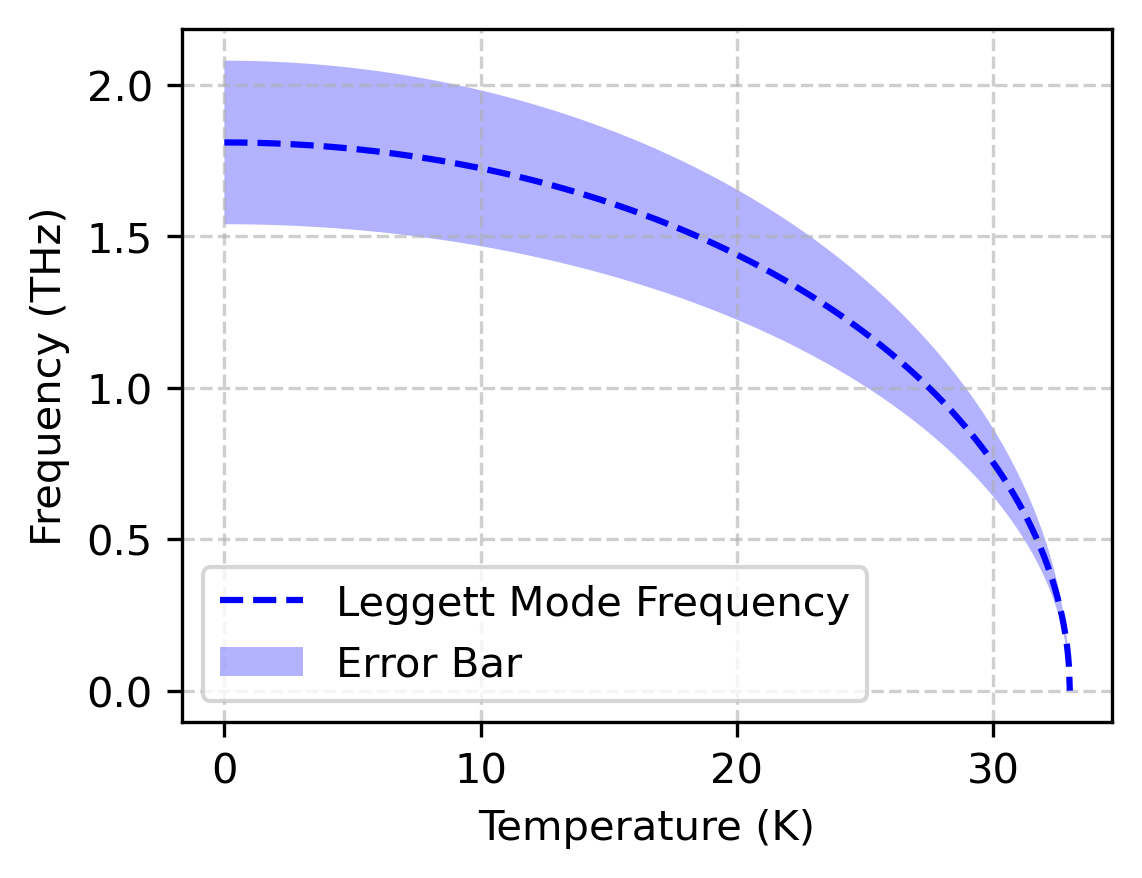}
        \renewcommand{\figurename}{Fig S}

    \caption{\textbf{Temperature dependence of calculated Leggett mode energy.}}    
    \label{Fig:S8}
\end{figure}

According to BCS theory, the superconducting energy gap $\mathrm{\Delta(T)}$ varies with temperature as:

\begin{equation}
\Delta(T) = \Delta(0) \sqrt{1 - \left(\frac{T}{T_c}\right)^2}
\end{equation}

Based on the relationship between $\mathrm{\omega_L}$ and $\mathrm{\Delta}$ in the above expression, the temperature dependence of $\mathrm{\omega_L}$ can be calculated, as shown in Fig.~S\ref{Fig:S8}.

\section{Comparison of Nonlinear signal and THz pump under Different Pump Fluences}

\begin{figure}[!h]
    \centering
    \renewcommand{\figurename}{Fig S}

    \begin{subfigure}[b]{0.5\linewidth}
        \centering
        \caption{Pump fluence: 13 kV/cm}
        \includegraphics[width=\linewidth]{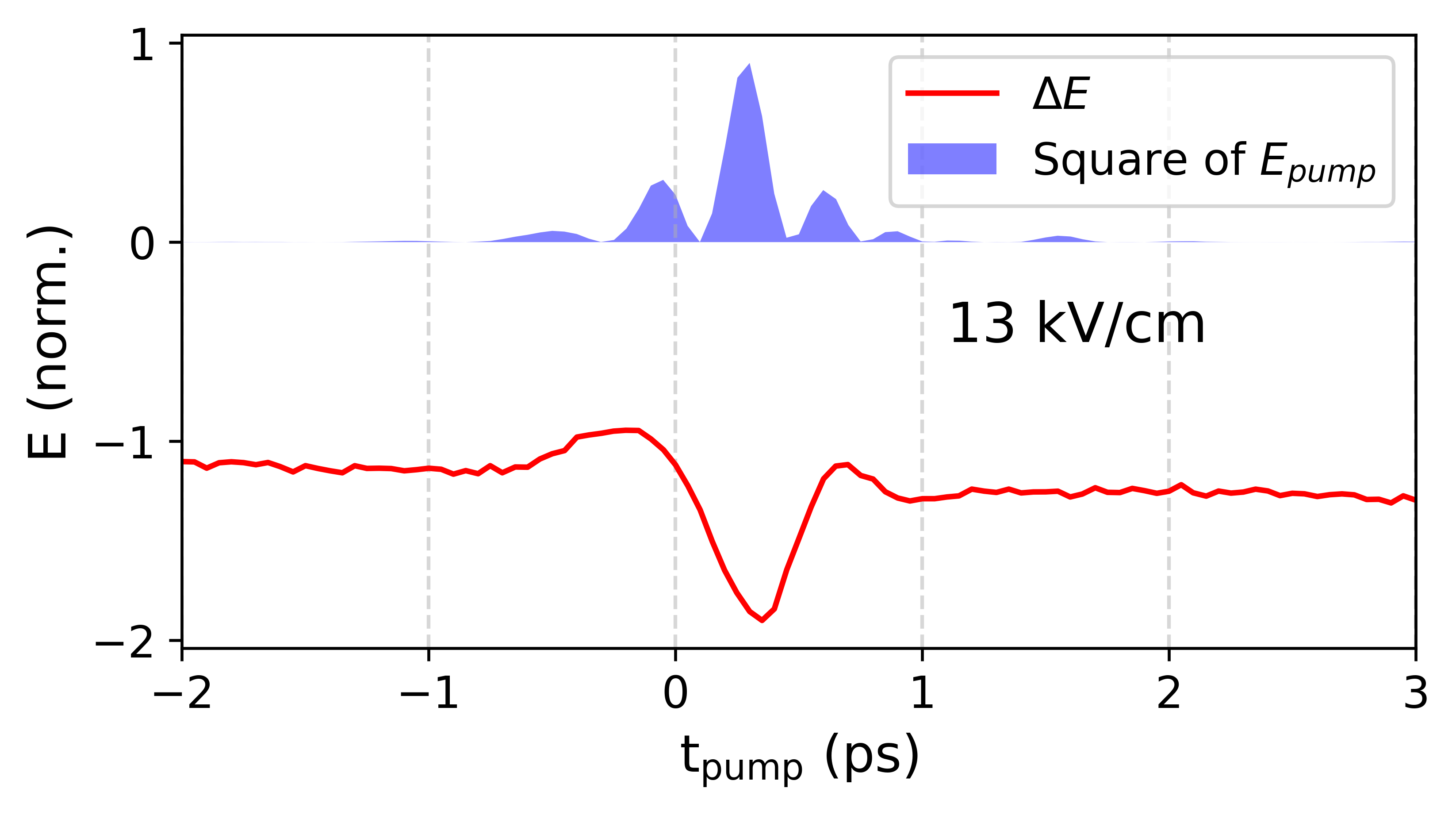}
        \label{fig:fluence_13}
    \end{subfigure}
    
    \begin{subfigure}[b]{0.5\linewidth}
        \centering
        \caption{Pump fluence: 35 kV/cm}
        \includegraphics[width=\linewidth]{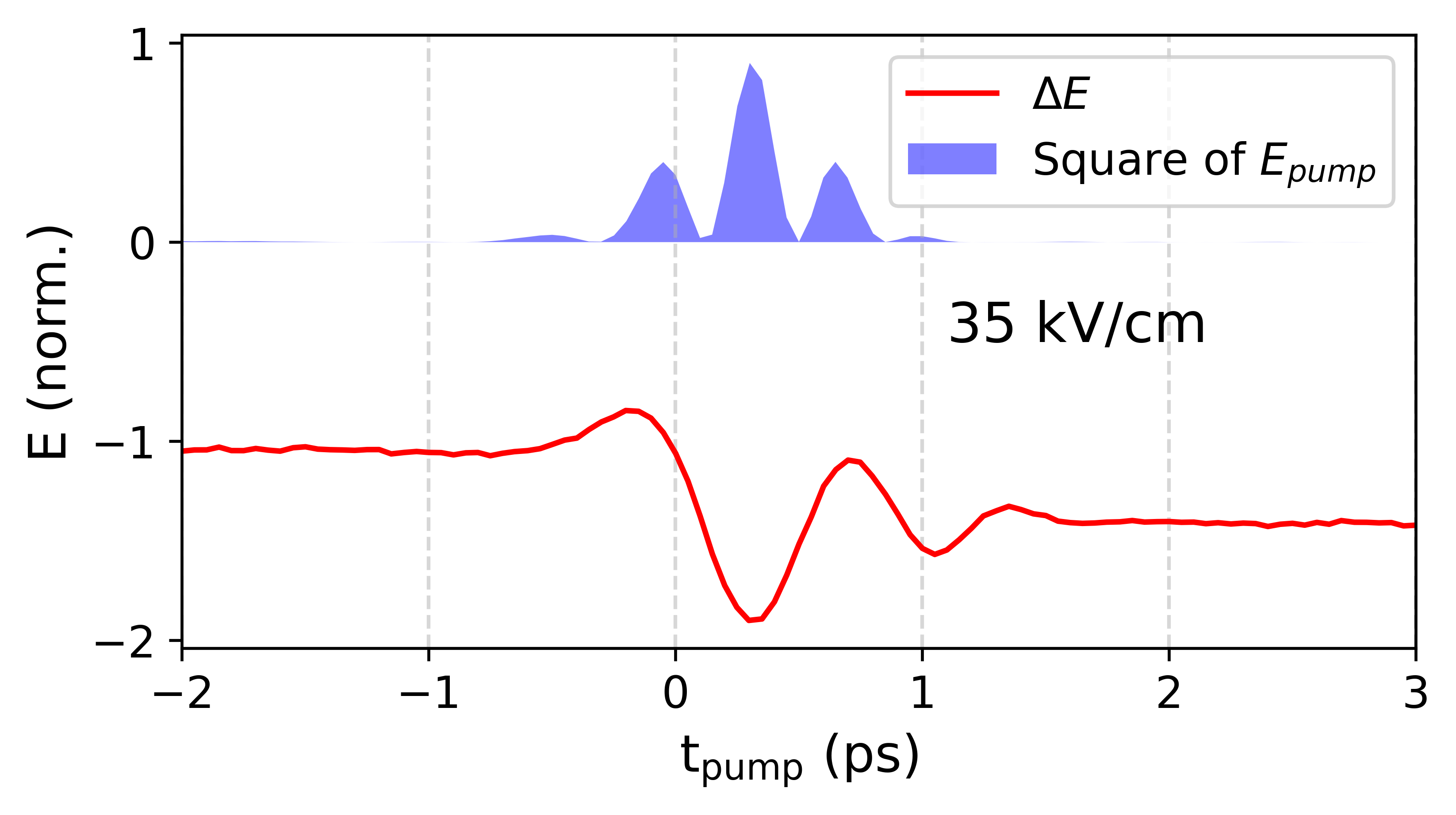}
        \label{fig:fluence_35}
    \end{subfigure}
    
    \begin{subfigure}[b]{0.5\linewidth}
        \centering
        \caption{Pump fluence: 250 kV/cm}
        \includegraphics[width=\linewidth]{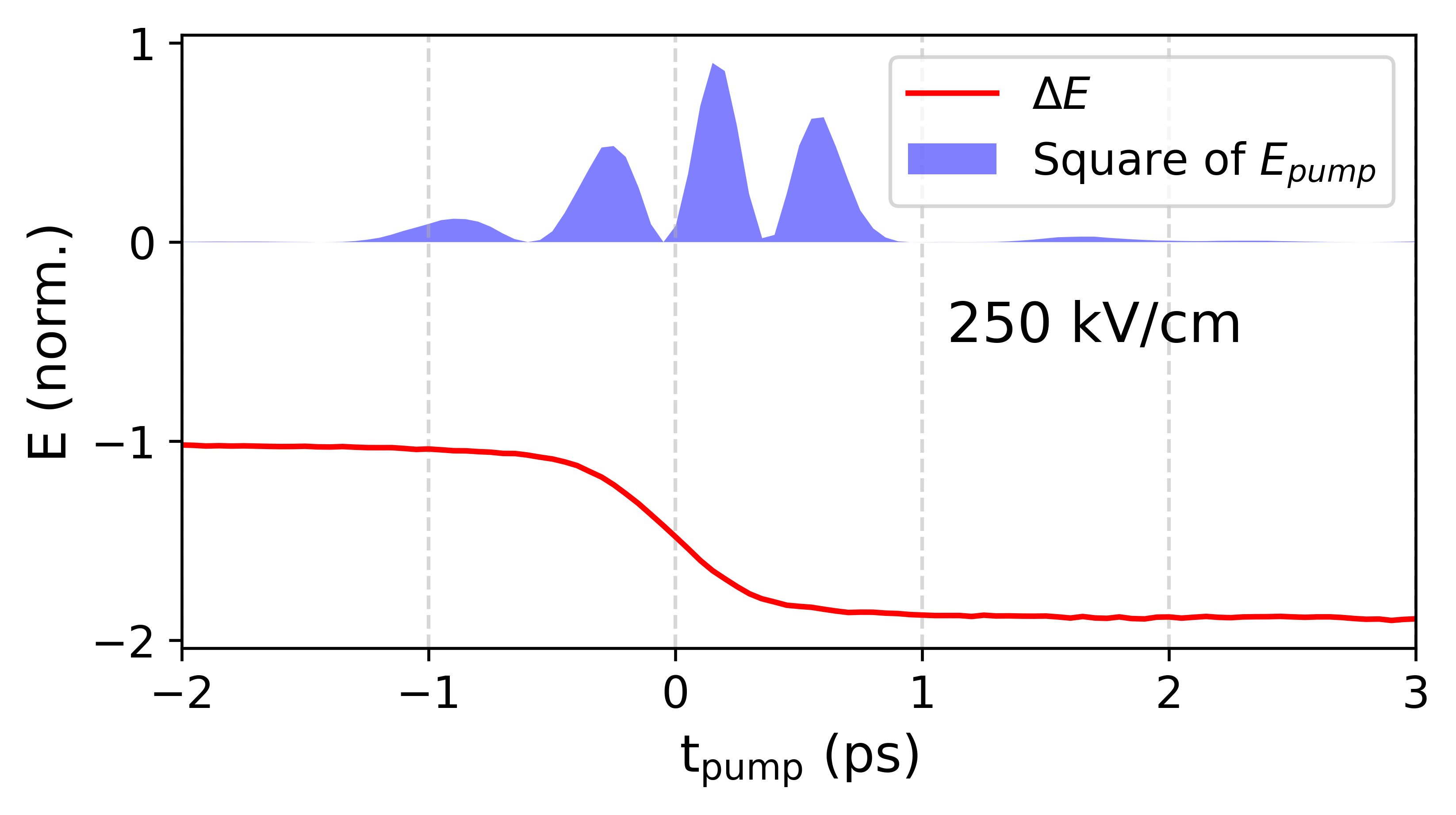}
        \label{fig:fluence_250}
    \end{subfigure}

    \caption{Normalized electric fields $\mathrm{E}$ as a function of time $t$ for different pump fluences (13, 35, and 250 kV/cm). Each panel shows the comparison between $\mathrm{\Delta E}$ and $\mathrm{E^2_{pump}}$.}
    \label{fig:fluence_vertical}
\end{figure}

To illustrate how the nonlinear signal detected in the single-cycle pump-probe experiment differs from the THz pump, we plotted the time-domain spectra of the nonlinear signal and the squared of the transmitted pump signal. The comparison reveals that at a pump fluence of 35 kV/cm (see Fig.~S\ref{fig:fluence_35}), the squared $\mathrm{E_{pump}}$ exhibits significantly higher frequencies in the time domain compared to the nonlinear signal. At 250 kV/cm (see Fig.~S\ref{fig:fluence_250}), the nonlinear signal behaves as a step function, which does not follow the behavior of the squared $\mathrm{E_{pump}}$. At 13 kV/cm (see Fig.~S\ref{fig:fluence_13}), the nonlinear signal nearly vanishes in the time-domain spectrum after 0.5 ps, where the pump still has significant spectral weight. Together, these observations indicate that the oscillatory behavior of the nonlinear signal has an optimal pump fluence (35 kV/cm) and further confirms that it does not originate from the THz pump.

\bibliography{ref3}
\end{document}